\documentclass[preprint,prb]{revtex4}
\usepackage{graphicx}
\usepackage{amsmath}
\usepackage{bbm}

\def\eq#1{Eq.~(\ref{#1})}
\def\fig#1{Fig.~\ref{#1}}

\def\sec#1{Sec.~\ref{#1}}

\newcommand{\boldv}[1]{{\mathbf #1}}
\newcommand{\ha}{\hspace{1cm}}

\newcommand{\de}{\mbox{d}}
\newcommand{\rez}[1]{\frac{1}{#1}}
\newcommand{\grad}{{\mathbf \nabla}}

\newcommand{\tr}{\mathrm{tr}}
\newcommand{\mean}[1]{\langle #1 \rangle}

\begin{document}

\title{Mean first passage times for bond formation for a Brownian particle in linear shear flow above a wall}
\author{C. B. Korn$^{1,2}$}
\author{U. S. Schwarz$^1$}
\affiliation{${}^1$University of Heidelberg, Im Neuenheimer Feld 293, D-69120 Heidelberg, Germany}
\affiliation{${}^2$Max Planck Institute of Colloids and Interfaces, D-14424 Potsdam, Germany}

\begin{abstract}
  Motivated by cell adhesion in hydrodynamic flow, here we
  study bond formation between a spherical Brownian particle in linear
  shear flow carrying receptors for ligands covering the boundary
  wall. We derive the appropriate Langevin equation which includes
  multiplicative noise due to position-dependent mobility functions
  resulting from the Stokes equation. We present a numerical scheme
  which allows to simulate it with high accuracy for all model
  parameters, including shear rate and three parameters describing
  receptor geometry (distance, size and height of the receptor
  patches). In the case of homogeneous coating, the mean first passage
  time problem can be solved exactly. In the case of position-resolved
  receptor-ligand binding, we identify different scaling regimes and
  discuss their biological relevance.
\end{abstract}

\maketitle

%\tableofcontents

\section{Introduction}

One of the hallmarks of biological systems is their tremendous
specificity in binding reactions between receptors and ligands. On the
molecular level, a prominent example is antigen-antibody recognition,
which allows our immune system to react to pathogens in a highly
specific way. Although traditionally much attention has been devoted
to the biochemical aspects of receptor-ligand binding, physical
concepts are equally important in this context. In particular, a
physical transport process is required to bring receptor and ligand to
sufficient proximity for binding. A helpful concept is the notion that
transport has to lead to the formation of an \textit{encounter
  complex}, which then can react to form the final receptor-ligand
complex \cite{r:eige74,c:berg77,r:shou82,c:schr02}.  In the language
of stochastic dynamics, the formation of the encounter complex is a
first passage problem which can be treated with appropriate tools from
statistical physics. In many situations, the transport process is
simple diffusion.  However, more complex situations also exist, like
the setup in affinity chips, where ligands are transported by
hydrodynamic flow into a reaction chamber loaded with receptors
\cite{r:wofs02}.

In cell adhesion, the physical transport processes required for
specific bond formation tend to be even more complex, because here
receptors and ligands are attached to surfaces and their movement is
determined by the dynamics of the objects they are attached to. One
important example in this context are white bloods cells, which
circulate the body with the blood flow and whose receptor-mediated
binding to ligand-coated walls is usually studied experimentally in
flow chambers \cite{c:spri94,c:alon95,c:chen01,uss:dwir03a}. In order
to fight pathogens in the surrounding tissue, white blood cells have
to extravasate from the blood vessels.  Initial binding is provided by
transmembrane receptors from the selectin family binding to
carbohydrate ligands on the vessel walls.  Here, the probability to
form an encounter complex is determined by the translational and
rotational movement of the cell as determined by hydrodynamic, thermal
and other external forces.  Similar situations also arise in
microbiology, when bacteria adhere to the intestinal wall
\cite{c:thom02}, in malaria infection, when infected red blood cells
adhere to the vessel walls \cite{c:bann03,c:naga00,c:amin05}, in the
initial stages of pregnancy, when the developing embryo adheres to the
uterus \cite{c:genb03}, and in biotechnology, e.\,g., when sorting
cells on microfluidic chips \cite{c:fore04}.

In this paper, we address this situation theoretically by combining
methods from hydrodynamics and stochastic dynamics. In
\fig{fig:cartoon} we show the situation which is theoretically
analyzed in the following. A spherical particle with radius $R$ moves
with hydrodynamic flow in positive $x$-direction at a height $z$ above
a wall. The simplest possible flow pattern is linear shear flow with
shear rate $\dot \gamma$. For the usual dimensions in flow chamber
experiments with white blood cells, this is the relevant flow profile.
In the absence of external forces, there is no reason for the particle
to drift towards the wall and the formation of an encounter complex
has to rely completely on thermal diffusion.  In many situations of
interest, however, there are forces pushing the particle towards the
wall, e.g., gravitational or electric forces.  In physiological blood
flow, cell density is high and the driving force for encounter is
provided mainly be hydrodynamic or contact interactions with other
cells. For the sake of computational simplicity and for conceptual
clarity, here we consider the simplest case of a force driving the
particle onto the wall, namely a constant gravitational force directed
in negative $z$-direction.  Therefore, we introduce a mass density
difference $\Delta \rho$ between the particle and the surrounding
fluid. Again this is the relevant situation in flow chamber
experiments, which are usually done with a diluted solution of cells,
thus ruling out a dominant role for cell-cell interactions.  Receptors
are modeled as patches on the particle surface, while ligands are
modeled as patches on the boundary wall. The formation of an encounter
complex is then identified with the first approach of any pair of
receptor and ligand patches which is smaller than a prescribed capture
radius $r_0$. The underlying stochastic process is rather complex due
to position-dependent mobilities resulting from the hydrodynamic
equations.

In order to solve the corresponding mean first passage time problem,
here we use computer simulations of the appropriate Langevin equation.
A short report of some of our results has been given before
\cite{uss:korn06a}. We start in \sec{sec:mobility} by introducing the
relevant concepts from hydrodynamics at small Reynolds numbers, in
particular the friction and mobility matrices resulting from the
Stokes equation for a rigid particle in linear shear flow above a
wall.  In \sec{sec:stokesian_dynamics} we combine these results with
concepts from stochastic dynamics in order to arrive at a Langevin
equation describing particle motion subject to hydrodynamic,
gravitational and thermal forces. Due to the position-dependent
mobility functions, we deal with multiplicative noise, that is special
care is needed to derive and interpret the noise terms.  In
\sec{sec:gravitation} our numerical scheme is applied to a sphere
falling in shear flow. The comparison of the measured stationary
height distribution function with the exact solution provides a
favorable test for our numerical treatment. In \sec{sec:MFPT} we show
that for the case of homogeneous coverage of sphere and wall the mean
first passage time to contact can be solved exactly, again in
excellent agreement with our numerical procedure. In
\sec{sec:initial_height} we explain why the choice for the initial
height is not essential. In the next two sections, we present and
explain our simulations results, first for movement restricted to two
dimensions in \sec{sec:2d} and then for the full three dimensional
case in \sec{sec:3d}.  We finally conclude in \sec{sec:discussion} by
discussing the biological and biotechnological relevance of our
results.

\section{Friction and mobility matrices}
\label{sec:mobility}

Due to their small sizes, the hydrodynamics of cells is in the low
Reynolds number regime. Using a typical cell size $L = 10\ \mu$m and a
typical velocity v = mm/s (that is the flow velocity at a distance $L$
to a wall with linear shear flow of rate $\dot \gamma =$ 100 Hz), the
Reynolds number is $Re = \rho v L / \eta = 10^{-2}$, where $\rho =$
g/cm$^3$ and $\eta = 10^{-3}$ Pa s are density and viscosity of water,
respectively. Therefore, we essentially deal with the Stokes equation
for incompressible fluids:
\begin{align}
  \label{linstokes}
  \eta \Delta \boldv{u}(\boldv{r}) - \boldv{\nabla}P(\boldv{r}) = 
 - \mathcal{{F}}(\boldv{r}), \ha \boldv{\nabla} \cdot \boldv{u}(\boldv{r}) = 0,
\end{align}
where $\boldv{u}(\boldv{r})$ is the fluid velocity field,
$P(\boldv{r})$ is the pressure field and $\mathcal{F}(\boldv{r})$ is
the force density on the fluid by the particle.  Here, we use
the induced force picture, i.e., the fluid equations of motion
\eq{linstokes} are extended to the interior of the particle and the
particle is replaced by an appropriate force density
$\mathcal{F}(\boldv{r})$ acting on the fluid \cite{felderhof:76}.
The unperturbed flow field has to satisfy the homogeneous version of
\eq{linstokes} as well as no-slip boundary conditions at the wall. In
this paper, we use the simplest possible example, namely linear shear
flow, $\boldv{u}^\infty = \dot \gamma z \boldv{e}_x$.

The effective flow field in the region occupied by the rigid sphere reads
\begin{align}
  \label{rigidbody}
  \boldv{u}(\boldv{r}) = \left(\boldv{U} + \boldv{\Omega}\times(\boldv{r} - \boldv{R})\right)
  \Theta(R - \|\boldv{r} - \boldv{R}\|)\ ,
\end{align}
where $\boldv{U},\boldv{\Omega}$ are the translational and rotational
velocities of the sphere, respectively. $\boldv{R}$ is the position of
its center, $R$ the sphere radius and $\Theta$ the theta
step-function. The particle is subject to forces and torques which
follow from the force density as
\begin{align}
  \label{forcemoments}
  \boldv{F}^H = \int\mathcal{F}(\boldv{r})\de\boldv{r}, \;\;
  \boldv{T}^H = \int(\boldv{r} - \boldv{R})\times\mathcal{F}(\boldv{r})\de\boldv{r}\ .
\end{align}
Because we consider a rigid object, higher moments of the force
density are not required in our context. For the unperturbed flow at
the mid-point of the sphere, we make the following definitions:
\begin{align}
  \label{velderiv}
         \boldv{U}^\infty =  \boldv{u}^\infty(\boldv{R}),\;\;
         \boldv{\Omega}^\infty = \rez{2} \left . \vphantom{\rez{2}}
  \boldv{\nabla}\times\boldv{u}^\infty(\boldv{r})\right|_{\boldv{r} = \boldv{R}},\;\; 
  \mathsf{E}^\infty_{ij} = \left . \vphantom{\rez{2}} 
\frac{1}{2} \left( \partial_i u^\infty_j(\boldv{r}) + \partial_j u^\infty_i(\boldv{r}) \right)
\right|_{\boldv{r} = \boldv{R}},
\end{align}
where the vector $\boldv{\Omega}^\infty$ is called \textit{vorticity} and the
tensor $\mathsf{E}^\infty$ \textit{rate of strain
  tensor}. Because we restrict ourselves to linear shear flow, all
higher moments of the unperturbed flow vanish.

Due to the linearity of the Stokes equation, a linear relationship
exists between the force density $\mathcal{F}(\boldv{r})$ and the
driving flow, which is the difference between real and unperturbed
flows \cite{felderhof:82}.  Specified for the first moments of the
force density, it leads to the relation
\begin{align}
  \label{frictionmatrix2}
  \left(\begin{array}{c}\boldv{F}^H\\\boldv{T}^H\end{array}\right) =
    -\mathsf{R_u}
  \left(\begin{array}{c}
    \boldv{U}^\infty - \boldv{U}\\
    \boldv{\Omega}^\infty - \boldv{\Omega}
    \end{array}\right) - \boldv{F}^S,
\end{align}
where the \textit{shear force} $\boldv{F}^S =
\mathsf{R_E}:\mathsf{E}^\infty$ with $\mathsf{A}:\mathsf{B} =
\tr~\mathsf{AB^T}$. It results from the perturbation of
the flow by the presence of the wall and vanishes for free flow. The
two matrices $\mathsf{R_u}$ and $\mathsf{R_E}$ are conveniently
written as
\begin{align}
  \label{friction}
  \mathsf{R_u} :=
    \left(\begin{array}{cc}
      \zeta^{tt}&\zeta^{tr}\\
      \zeta^{rt}&\zeta^{rr}
    \end{array}\right),\ha
    \mathsf{R_E} := 
    \left(\begin{array}{c}
      \zeta^{td}\\
      \zeta^{rd}
    \end{array}\right),\ha
\end{align}
where the $\zeta$ are the symmetric \textit{friction matrices} and the
superscripts t, r and d stand for \textit{translational},
\textit{rotational} and \textit{dipolar}, respectively.  In order to
obtain the translational and rotational velocities of the sphere as a
function of the hydrodynamic forces and torques, we have to invert
\eq{frictionmatrix2}:
\begin{align}
  \label{motion01}
  \left(\begin{array}{c}\boldv{U}\\\boldv{\Omega}\end{array}\right) =
  \left(\begin{array}{c}
    \boldv{U}^\infty\\
    \boldv{\Omega}^\infty\end{array}\right)
  + \mathsf{M} \left[ \left(\begin{array}{c}\boldv{F}^H\\\boldv{T}^H\end{array}\right)
  + \boldv{F}^S \right]\ .
\end{align}
The symmetric matrix $\mathsf{M} = \mathsf{R_u}^{-1}$ is called
\textit{mobility matrix}. It is convenient to define the mobility
tensors through
\begin{align}
  \label{mobilitymatrix}
  \mathsf{M} = \mathsf{R_u}^{-1} = 
  \left(\begin{array}{cc}
    \mu^{tt}&\mu^{tr}\\
    \mu^{rt}&\mu^{rr}
  \end{array}\right), \ha
    \mathsf{R_u}^{-1}\mathsf{R_E} = \left(\begin{array}{c}
    \mu^{td}\\
    \mu^{rd}
  \end{array}\right)\ .
\end{align}
In order to calculate the friction and mobility tensors for the
special case of a sphere in linear shear flow above a wall, we follow
the procedure from Ref.~\cite{jones:98}. The friction tensors $\zeta$
introduced in \eq{friction} and the mobility tensors $\mu$ introduced
in \eq{mobilitymatrix} are expressed in terms of scalar functions
together with irreducible tensors formed form the Kronecker symbol
$\delta_{ij}$, the Levi-Civita symbol $\epsilon_{ijk}$ and the normal
vector $\boldv{k} = \boldv{e}_z$. The scalar friction and mobility
functions are not known analytically, but can be obtained to high
accuracy by the following numerical scheme. One introduces the
variable $t = R/z$, where $R$ is the radius of the sphere and $z$ is
its height above the wall. Thus, $t$ can take values from the interval
$[0,1]$. In the limit $t \to 0$, that is far away from the wall, one
can expand the friction functions in powers of $t$.  In the limit $t
\to 1$, that is close to the wall, analytical results can be obtained
with lubrication theory. In order to cover the whole interval, the two
limit solutions are matched using a Pad\'e summation scheme \cite{jones:98}.  More
details of this implementation are given in appendix
\ref{appendix:mobility}. In contrast to the tabulated finite element
results from Ref.~\cite{goldman:67b}, this implementation gives correct
results for any possible configuration.

\section{Langevin equation}
\label{sec:stokesian_dynamics}

The motion of a particle subject to thermal, hydrodynamic and direct
external forces like gravity is called \textit{Stokesian Dynamics}
\cite{brady:89}. In this section we derive the corresponding
stochastic differential equation (\textit{Langevin equation}). The
Langevin equation will allow us to base our statistical treatment on
the repeated simulation of individual trajectories. Because we are
interested in the over-damped (Stokes) limit, we can neglect inertia
in Newton's second law:
\begin{align}
  \label{allforces}
  -\boldv{F}^H + \boldv{F}^D + \boldv{F}^B = 0,
\end{align}
where $- \boldv{F}^H$, $\boldv{F}^D$ and $\boldv{F}^B$ are hydrodynamic,
direct and thermal forces acting on the sphere. An analogous balance
exists for the torques. For the following, forces and torques as
described above are united in one symbol. For example, from now on the
symbol $\boldv{F}$ denotes $(\boldv{F},\boldv{T})$, a six-dimensional
vector comprising force $\boldv{F}$ and torque $\boldv{T}$, and
$\boldv{U}$ denotes the six-dimensional
particle translational/rotational velocity vector. \\
In the absence of Brownian forces, $\boldv{F}^B = 0$ and $\boldv{F}^D
= \boldv{F}^H$.  Inserting this into \eq{motion01} then gives
\begin{align}
  \label{det-algorithm}
  \boldv{U} = \boldv{U}^\infty + \mathsf{M}(\boldv{F}^D +
  \boldv{F}^S)
\end{align}
and the particle trajectory can be found with a simple Euler algorithm
as $\boldv{X}(t + \Delta t) = \boldv{X}(t) +
\boldv{U}\Delta t + \mathcal{O}(\Delta t^2)$.

In the presence of Brownian motion, the situation is more complex,
because thermal noise leads to terms of the order $\Delta t^{1/2}$
and special care has to be taken to include all terms up to order $\Delta t$.
Due to the fluctuation-dissipation theorem, for our problem
Gaussian white noise reads
\begin{align}
  \label{noise}
  \mean{\boldv{g}_t} = 0, \ha \mean{\boldv{g}_t \boldv{g}_{t'}} = 
  2k_BT \mathsf{M}\delta(t - t')\ .
\end{align}
Here, the subscript $t$ corresponds to the fact that the thermal force
$\boldv{g}$ is a random process.  The left part of \eq{noise} states
that the forces that the fluid exerts on the particles are equally
distributed in all directions so there is no net drift due to thermal
fluctuations.  The right part of \eq{noise} states that forces at
different times are not correlated, which is a good approximation
because the diffusive forces act on a much faster time scale than the
hydrodynamic forces. Because the mobility matrix $\mathsf{M}$ is
position-dependent, we deal with so-called \textit{multiplicative
  noise}. Since the $\delta$-correlation in \eq{noise} can be
considered to be the limit of a process with an intrinsic time scale
for thermal relaxation, which is much faster than the time scale of
hydrodynamic movement, the Stratonovich interpretation of the
stochastic process is appropriate \cite{vankampen:92}. This means
that for each time step, the mobility functions have to be evaluated
at $\boldv{X}(t + (1/2)\Delta t)$ (rather than at $\boldv{X}(t)$ as in
the It\^o interpretation).  The Stratonovich interpretation also
implies that the rules for integration and coordinate transformation
are the same as for the Riemann integral in non-stochastic calculus.

The presence of the thermal noise \eq{noise} converts the position
function $\boldv{X}(t)$ into a random process $\boldv{X}_t$. 
Multiplicative noise can result in additional drift terms. We
therefore write the Langevin equation as
\begin{align}
  \label{langevin-01}
  \partial_t \boldv{X}_t = \boldv{U}^\infty + \mathsf{M}(\boldv{F}^D +
  \boldv{F}^S) + k_BT \boldv{Y} + \boldv{g}^S_t, 
\end{align}
where in comparison to the deterministic equation \eq{det-algorithm}
we have added both the Gaussian white noise $\boldv{g}^S_t$ (to be
interpreted in the Stratonovich sense) and some drift term
$\boldv{Y}$. The drift term $\boldv{Y}$ can be derived by requiring
\eq{langevin-01} to be equivalent to the appropriate Smoluchowski
equation. The details of these calculations are given in appendix
\ref{appendix:gradient_term}. The result is
\begin{align}
\label{matrixB}
  \boldv{Y} = \mathsf{B}\grad\mathsf{B}^T,\ha
  \mathsf{M} = \mathsf{BB}^T,\ha
  Y_i = \mathsf{B}_{ik}(\partial_l\mathsf{B}_{lk}),\ha
  \mathsf{M}_{ij} = \mathsf{B}_{ik}\mathsf{B}_{jk}\ .
\end{align}
For additive noise, that is for position-independent mobility
functions, the additional drift term would vanish. In the case of
position-dependent mobility matrices, the noise term $\boldv{g}^S_t$
alone would lead to a drift of the particle towards regions of lower
mobility (that is towards the wall, where mobility vanishes due to the
no-slip boundary condition).  This drift, however, is exactly
compensated by the additional term $\boldv{Y}$.

For the following, it is useful to non-dimensionalize \eq{langevin-01}.
For length, the natural scale is sphere radius $R$. For time, we use
$6\pi\eta R^3/k_BT$, which is the time needed to diffuse the distance
$R$. For force, we use $6\pi\eta R^2\dot\gamma$, the Stokes force at
velocity $R \dot\gamma$, that is in linear shear flow a distance $R$ away
from the wall. The scalar friction and mobility functions appearing in
$\mathsf{M}$, $\mathsf{R_E}$ and $\mathsf{R_u}$, also become
dimensionless as explained in appendix \ref{appendix:mobility}.  The
Langevin equation \eq{langevin-01} now reads
\begin{align}
  \label{langevin-04}
  \partial_t \boldv{X}_t = Pe\left(\boldv{U}^\infty +
  \mathsf{M}(f \boldv{F}^D + \boldv{F}^S )\right)
  + \mathsf{B}\grad\mathsf{B}^T + \boldv{g}^S_t,
\end{align}
where the \textit{P\'eclet number} $Pe = 6\pi\eta R^3\dot\gamma /
k_BT$ measures the relative importance of deterministic to Brownian
motion. In the limit $Pe \rightarrow 0$ the particle only exhibits
diffusive motion and in the limit $Pe \rightarrow \infty$ it is no
longer subjected to diffusion. The second dimensionless parameter $f =
\|\boldv{F}^D\| / 6\pi\eta R^2\dot\gamma$ measures the relative
importance of direct forces/torques versus the shear force/torque.
Measuring the time in units of the diffusive time scale is appropriate
for P\'eclet numbers of order ten or less. For simulations with larger
P\'eclet numbers it is more suitable to scale time with the inverse
shear rate $\dot\gamma^{-1}$. This has the effect of dividing
\eq{langevin-04} by $Pe$.

In order to solve \eq{langevin-04} numerically, it has to be
discretized with respect to time.  The appropriate Euler algorithm can
be derived by first rewriting \eq{langevin-04} in the It\^o-version,
which adds another drift term to the equation.  As explained in
appendix \ref{appendix:euler}, the two drift terms together lead to
the result
\begin{align}
  \label{langevin-ito}
  \partial_t \boldv{X} = Pe\left(\boldv{U}^\infty +
  \mathsf{M}( f \boldv{F}^D + \boldv{F}^S )\right)
  + \grad\mathsf{M} + \boldv{g}^{I}_t\ .
\end{align}
Its discretized version is simply
\begin{equation}
  \label{langevin-euler}
  \Delta \boldv{X} = \left [\left . Pe\left(\boldv{U}^\infty + 
  \mathsf{M}( f \boldv{F}^D + \boldv{F}^S )\right)\right|_t
  + \left .\grad\mathsf{M}\right|_t\right]\Delta t + \boldv{g}(\Delta t) + \mathcal{O}(\Delta t^2)\ .
\end{equation}
This final result has been derived before in a different way by Brady
and Bossis \cite{brady:89}. For vanishing shear flow, it also agrees
with the classical result by Ermak and McCammon \cite{ermak:78}. In
appendix \ref{appendix:euler}, we describe the algorithms used to
implement \eq{langevin-euler}, in particular the algorithm to generate
the thermal forces $\boldv{g}(\Delta t)$.

\section{Sphere falling in shear flow}
\label{sec:gravitation}

As explained in the introduction, we consider a sphere whose density
is slightly larger than that of the fluid. Due to this density
difference $\Delta \rho$ a constant drift towards the wall exists. As
we will see later, this drift ensures that on average the sphere will
bind to the wall in finite time.  The two independent parameters
defined in \eq{langevin-04} for this model system are $Pe$ and $f = (2
R\Delta \rho g)/(9 \eta\dot\gamma)$, with the earth acceleration
constant $g = 9.81$ m/s$^2$.  For later considerations, it is
convenient to introduce also the parameter $Pe_z = f\ Pe$, which we
call the \textit{P\'eclet number in $z$-direction}.  $Pe$ and $Pe_z$
represent the strengths of the hydrodynamic and gravitational forces
in respect to the thermal force, respectively. Out of the three
parameters $Pe$, $f$ and $Pe_z$, only two are independent, because $f
= Pe_z / Pe$. 

We first consider the path of a sphere falling in shear flow after it
has been dropped at some initial height.
\fig{falling} illustrates the effect of the P\'eclet number by showing
some representative simulation trajectories. For $Pe = \infty$ the
motion of the sphere is purely deterministic and only governed by the
parameter $f$.  In the diffusive limit $Pe = 0$, the sphere makes a
pure random walk (except for the drift in $z$-direction due to the
gravitational force).

As the mobility matrix does only depend on the height of the sphere
above the wall (cf. appendix~\ref{appendix:mobility}), the motion in
the $z$-direction is independent of the position in the $(x,y)$-plane
and the orientation of the sphere. Therefore, it can be treated
separately.  The probability density $\Psi(z,t)$ for the sphere to be
at height $z$ at time $t$ is the solution to a one dimensional
Smoluchowski equation
\begin{align}
    \label{smol-z}
  \partial_t\Psi(z,t) = - \partial_z J_z,\
  J_z = - \mathsf{M}_{zz}(\partial_z \Psi + Pe_z \Psi).
\end{align}
This equation cannot be solved analytically as the mobility
function $M_{zz}$ is not known in closed form. In \fig{fig-z-dist} we show
numerical solutions obtained by simulating the equivalent Langevin
equation.
One clearly sees that first the $\delta$-function at $t = 0$ is
broadened due to diffusion and then develops into a stationary solution
which has its maximum at the wall. This stationary solution has a
simple analytical form which follows from \eq{smol-z} by integrating
$J_z = 0$:
\begin{align}
  \label{statsol}
  \Psi_s(z) = Pe_z e^{-Pe_z (z - 1)}\ .
\end{align}
Thus, the stationary solution is simply the barometric formula, as it
should be for thermodynamic reasons. We also find that the first two
moments (mean and variance) are the same:
\begin{align}
  \mean{z - 1} = \sqrt{\mean{z^2}-\mean{z}^2} = \rez{Pe_z}\ .
\end{align}
In the limit of vanishing gravitational force ($Pe_z \rightarrow 0$),
the probability distribution becomes flat and the probability of
finding the sphere does not peak at the wall anymore.

\section{First contact with homogeneous coverage}
\label{sec:MFPT}

If the sphere and the wall are homogeneously covered with receptors
and ligands, respectively, an encounter complex is established
whenever the sphere comes sufficiently close to the wall. The mean
time which elapses after the sphere is set free at some initial
position until an encounter complex is established is then identical
with the mean first passage time (MFPT) for a sphere dropped at initial
height $z_0$ to reach the height $z_1$. Note again that the motion in
$z$-direction is independent of the values of the other coordinates.
For a particle diffusing in an interval $[z_1,b]$, with $z_1$ being an
absorbing boundary and $b$ a reflective boundary, the MFPT
$T$ to reach $z_1$ when started at $z \in [z_1, b]$ is
the solution to the following ordinary differential equation
\cite{vankampen:92}
\begin{align}
  \label{fpt-02}
  A(z)\partial_z T(z|z_1) + D(z)\partial^2_z T(z|z_1) = -1, \ha
  T(z_1|z_1) = 0,\ha \left.\partial_z T(z|z_1)\right|_{z = b} = 0.
\end{align}
In our case, $b = \infty$. The drift term is $A(z) =
-Pe_z\hat\alpha^{tt}(1/z) + \partial_z\hat\alpha^{tt}(1/z)$ and the
diffusive term $D(z) = \mathsf{M}_{zz} = \hat\alpha^{tt}(1/z)$,
where $\hat\alpha^{tt}(1/z)$ is a scalar mobility function as
explained in appendix \ref{appendix:mobility}.  The general solution
to \eq{fpt-02} is \cite{vankampen:92}
\begin{align}
  \label{mfpt-general}
   T(z_0| z_1) = \int\limits_{z_1}^{z_0}\de z \rez{\Phi(z)}
   \left(\int\limits_z^\infty \de y
  \frac{\Phi(y)}{D(y)}\right), \ha
  \Phi(z) = \exp\left(\int\limits^z \de x \frac{A(x)}{D(x)}\right).
\end{align} 
This, can be reduced up to an integral over $\hat \alpha^{tt}(1/z)$:
\begin{align}
  \label{z-mfpt}
  T(z_0| z_1) = \rez{Pe_z}\int\limits_{z_1}^{z_0}\de z \rez{\hat \alpha^{tt}(1/z)}.
\end{align}
Thus the dependence of $T(z_0| z_1)$ on $Pe_z$, the only parameter
in this problem, is obtained exactly. It is important to note that the
compact form for the MFPT in \eq{z-mfpt} is a result of the constant
vertical force.  For a more general vertical potential force $F_\perp
= -\partial_z V(z)$ with a potential $V$, \eq{mfpt-general} can be
reduced to
\begin{align}
  \label{mfpt-potential}
  T(z_0|z_1) = \int_{z_1}^{z_0}\frac{dz}{\hat\alpha^{tt}(1/z)}
  \int_z^\infty d y e^{V(z) - V(y)}\ .
\end{align}
This equation shows that the potential must satisfy the condition
$\lim_{y\rightarrow\infty}(V(z) - V(y))\rightarrow -\infty$ for the
MFPT to be finite. This holds true, e.g., for the gravitational force
studied here or for the interaction of a charged object with an
oppositely charged wall, but not, e.g., for a Lennard-Jones potential.

The integral \eq{z-mfpt} over the scalar
mobility function $\hat\alpha^{tt}$ can easily be calculated
numerically as $\hat\alpha^{tt}$ behaves well in the full range of
$z$.  In fact $\hat\alpha^{tt}(t)$ can be approximated by its leading
term from the lubrication analysis, i.\,e.,  $\hat\alpha^{tt}(t)
\approx 1 - t$. We then find
\begin{align} \label{mfpt:approx}
  T(z_0| z_1) \approx  \rez{Pe_z}\left[z_0 - z_1 + 
    \ln\left(\frac{z_0 - 1}{z_1 - 1}\right)\right].
\end{align}
A numerical analysis shows that the approximation \eq{mfpt:approx}
deviates only by a few percent from the exact solution \eq{z-mfpt}.
Thus, $T(z_0| z_1)$ is logarithmically divergent if the absorbing
point is close to the wall, $z_1 \rightarrow 1$, and linearly
divergent if the starting point is at infinite height, $z_0
\rightarrow \infty$.

For a sphere homogeneously covered with receptors each having a
capture radius $r_0$, the mean time for forming an encounter complex
is $T(z_0| 1 + r_0)$.  This time will serve as a useful limiting
result in some of the considerations presented in the next sections.
The exactly known result \eq{z-mfpt} provides also a good test for the
algorithm we implemented. In \fig{mfpt}a the MFPT
obtained from simulation experiments and from quadrature of
\eq{z-mfpt} are compared. The two results agree very well (see
appendix \ref{appendix:error} for a discussion of the statistical and
systematic errors of the simulation results). In \fig{mfpt}b we show
the numerically obtained distribution of first passage times. One
clearly sees that the larger $Pe_z$, the stronger they peak around the
mean.

We conclude the case of homogenous coverage by noting that in order to
obtain dimensionalized results, one has to multiply the MFPT by
the diffusive time scale $6\pi\eta R^3 / k_B T$. This result does not
depend on shear rate $\dot \gamma$ because vertical and horizontal
motion are decoupled and rotational motion is not relevant
here. However, it depends on viscosity $\eta$, which sets the time
scale for vertical motion. If one switched off thermal fluctuations,
the falling time would be exactly the same as the MFPT
from \eq{z-mfpt}, but this is a special result for constant
force and not true in general. If one removed the wall, the
translational symmetry in $z$-direction would not be broken and the
MFPT would be $T = (z_0 - z_1)/Pe_z$, that is the
logarithmic term in \eq{mfpt:approx} would be missing.

\section{Effect of initial height}
\label{sec:initial_height}

We now turn to spatially resolved receptor coverage, that is we
consider a sphere which is covered by $N_r$ equidistantly spaced
receptor patches. For the moment being, the wall is still considered
to be homogeneously covered with ligands. The MFPT
$T(\vec{\theta}, \vec{x}| C)$ now will depend on the initial
position $\vec{x} = (x, y, z_0)$ and the initial orientation
$\vec{\theta}$ as well as on the absorbing boundary $C$ in diffusion
space. The latter is given by the special receptor and ligand
geometry.  In an experimental setup with linear shear flow it is
possible to measure only particles which have been initially at a certain
height.  This is due to the fact that their average velocity as
obtained from the solution of the Stokes equation \eq{motion01}
depends on their height in a unique way \cite{c:chen01}. However, it
is almost impossible to prepare a certain initial orientation
$\vec{\theta}$ or $(x,y)$-position relative to the ligands.  Therefore,
the quantity of interest to us will be a MFPT which
is averaged over all possible initial orientations $\vec{\theta}$ and
all initial positions $(x,y)$, which will be denoted as
$\mean{T(\vec{\theta}, \vec{x} | C)}_{\vec{\theta}, (x,y)}$.
The dependence of $\mean{T(\vec{x}, \vec{\theta}|
  C)}_{\vec{\theta}, (x,y)}$ on the initial height for $z_0 > 1 + r_0$
can be derived exactly. For homogeneous ligand coverage the quantity
of interest is
\begin{align*}
  \mean{T(\vec{\theta}, z_0 | C)}_{\vec{\theta}} = 
  \rez{V_{\vec{\theta}}}\int_{\vec{\theta}}d^3\vec{\theta}\;
  T(\vec{\theta}, z_0 | C),
\end{align*}
where $C$ is the absorbing hyper-surface in $(\vec{\theta}, z)$-space and
$V_{\vec{\theta}}$ a normalization constant. Absorption is only
possible if $z < 1 + r_0$, thus if we look at some intermediate height
$z_0 > z_m > 1 + r_0$, then
\begin{align}
  \label{zdep:two}
  T(\vec{\theta}, z_0 | C) = T(\vec{\theta}, z_0 | z_m) + \int d^3
  \vec{\theta}_m\; p(\vec{\theta}_m | \vec{\theta})T(\vec{\theta}_m, z_m | C),
\end{align} 
where $p(\vec{\theta}_m | \vec{\theta}_m)$ is the conditional probability
to pass the height $z_m$ with the orientation $\vec{\theta}_m$ when
starting with the initial orientation $\vec{\theta}$ at
$z_0$. $T(\vec{\theta}, z_0 | z_m)$ is independent of
the initial orientation and can be calculated by means of
\eq{z-mfpt}. Now averaging \eq{zdep:two} over the initial orientation
gives
\begin{align}
  \label{mfpt:add}
  \mean{T(\vec{\theta}, z_0 | C)}_{\vec{\theta}_m} &= T(z_0 | z_m) + 
  \rez{V_{\vec{\theta}}}\int d^3
  \vec{\theta}_m\; \left[\int d^3
  \vec{\theta}\;p(\vec{\theta}_m | \vec{\theta})\right]T(\vec{\theta}_m,
  z_m | C)\\ \nonumber
  &= T(z_0 | z_m) + \rez{V_{\vec{\theta}}}\int d^3
  \vec{\theta}_m\; T(\vec{\theta}_m,
  z_m | C) = T(z_0 | z_m) + \mean{T(\vec{\theta}_m, z_m | C)}_{\vec{\theta}_m}. 
\end{align} 
Thus, if the orientation-averaged MFPT is known for
some initial height $z_0 > 1 + r_0$, then the MFPT
for any other initial height $z_0' > 1 + r_0$ can be calculated by
means of equations \eq{mfpt:add} and \eq{z-mfpt}. In
\fig{initial_height}, this result is verified by simulations for the
two-dimensional case, that is the sphere can only move in the
$x-z$-plane and rotate only around the $y$-axis, compare
\fig{fig:cartoon}. Due to the decomposition \eq{mfpt:add}, the initial
height is not essential.  In the following, we therefore will always
use the value $z_0 = 2$, that is the sphere has to fall by one radius
until it hits the substrate for the first time.

\section{Movement in two dimensions}
\label{sec:2d}

We now study the effect of shear rate for heterogeneous receptor
distribution if the sphere is restricted to move only in two
dimensions. Then, the receptor patches can be equidistantly distributed
over the circumference as illustrated in \fig{fig:rpatch}.  Each
receptor patch has a capture height of $r_0$ and a width of $2r_p$.
The 2D receptor density is then $\rho_r = N_r r_p / \pi$.  Orientation
is now represented by a single angle $\theta$.  The absorbing boundary
$C$ is illustrated in \fig{fig:rpatch}. For each receptor patch,
binding can occur over a range $2 \theta_0$, which consists of two
parts. The inner part is valid already for $r_p = 0$ and reflects the
overlap due to a finite $r_0$. The outer part is results from a finite
$r_p$.  Together this leads to $\theta_0(z) = \arccos(z/(1 + r_0))+
r_p$.  The receptor patches establish a periodicity with period
$\theta_s = 2\pi/N_r$.  As the number of receptor patches grows, this
period decreases and one finally achieves overlap. Then, encounter
becomes possible for all values of $\theta$, that is we are back to
the case of homogeneous receptor coverage. In our case of
non-homogeneous coverage, the MFPT depends on $Pe, Pe_z, N_r, r_0,
r_p$ and $z_0$.  For the following simulations $r_p = r_0 = 10^{-3}$,
$Pe_z = 50$ and $z_0 = 2$ is chosen unless other values are explicitly
mentioned.

\fig{fig:receptorresults:one}a shows the MFPT as a
function of the P\'eclet number $Pe$. Note that in the
log-log plot, an apparent plateau appears at small
value of $Pe$, although in a linear plot there
would be monotonous decay. Three regimes can be
distinguished.  For $Pe \approx 0$ (\textit{diffusive limit}) the
transport by the imposed shear flow is negligible and only diffusive
transport is present. For very large values of $Pe$,
$\mean{T}_\theta$ plateaus at the value given by \eq{z-mfpt}
independent of $N_r$. In this limit the time for rotation to any
certain orientation is negligible compared to the mean time to fall
down close to the wall, therefore, the result for rotational symmetry
is recovered. Between these two limits the MFPT
decreases monotonically with increasing $Pe$.
\fig{fig:receptorresults:one}b shows the data from
\fig{fig:receptorresults:one}a plotted as a function of the receptor
density $\rho_r \propto N_r$. The larger $Pe$ the less pronounced is
the dependence of $\mean{T}_\theta$ on $N_r$. For $Pe \approx 0$,
however, $\mean{T}_\theta$ strongly depends on $N_r$.  The latter
relation is better illustrated in \fig{fig:receptorresults:one}c.
There, at $Pe \approx 0$, $\mean{T}_\theta$ is shown for a wide
range of $N_r$. The simulations were done for fixed patch size $r_p$
but for four different values of the capture radius $r_0$ (cf.
\fig{fig:rpatch}).  For $\rho_r \rightarrow 1$, $\mean{T}_\theta$
reaches the value given by \eq{z-mfpt}. As described by \eq{z-mfpt},
$\mean{T}_\theta$ is the smaller the larger $r_0$ is. An increase
in the number of receptor patches $N_r$ leads to a strong decrease for
the MFPT, however, no special scaling behavior can
be observed.  It is remarkable that the limiting value for the case of
homogeneous receptor coverage is already reached for $\rho_r \approx
10^{-2}$. The larger the capture radius $r_0$ the more pronounced is
this effect. This can be understood by observing that the effective
patch size as given by the angle $\theta_0 \geq r_p$ (see
\fig{fig:rpatch}) is monotonically increasing with increasing $r_0$.

We next try to qualitatively understand the effect of shear rate for
the simulation results shown in \fig{fig:receptorresults:one}a. In
general, it is very hard to separate the effects of diffusion and
convection. The time for binding at $Pe \approx 0$ is determined
purely by diffusion effects and will be denoted by $T_D$. As shear
flow increases, the rotation of the sphere is increasingly dominated
by convection.  We now derive a convection time $T_F$ which competes
with the diffusion time $T_D$ at large P\'eclet number. For very large
P\'eclet number, we expect the MFPT to be the sum of the homogeneous
result from \eq{mfpt:approx} plus this additional time $T_F$. An
important question then is at which $Pe$ the convection time $T_F$
become smaller than the diffusion time $T_D$.

On order to estimate $T_F$, we note that the main effect of increased
shear rate is faster rotation in the direction of flow.  Once a
receptor has rotated by an angle $\theta_s = 2\pi/N_r$ such that it
opposes a ligand on the substrate, there is some probability $p$
that the sphere is at the correct height that an encounter can occur.
If no encounter occurs with the complementary probability $1-p$, the
sphere has to rotate about another angle $\theta_s$ until the next
receptor points downwards.  Supposing the time, $2t_0$, to
rotate about the angle $\theta_s$ is large enough that there is no
correlation between the height of the sphere before and after the
rotation, then, an encounter occurs again with probability $p$
(therefore this analysis also does not hold at very large $Pe$). Thus,
the mean time $T_F$ for encounter is
\begin{align}
  \label{mfpt:flowtime:one}
  T_F &= p t_0 + (1 - p)(p3t_0 + (1 - p)(p5t_0 + (1-p)( \ldots))) \nonumber \\
  &= p t_0 \sum_{i = 0}^\infty (2i + 1)(1 - p)^i = t_0 \frac{2-p}{p}
   \approx \frac{2 t_0}{p}\ ,
\end{align}
where the series has been summed up by means of the geometric
formula. In the last term we assumed that the probability $p$ for the
proper height is small due to a small capture distance $r_0$. It
follows from the stationary probability distribution $\Psi_s(z)$ given
by \eq{statsol}:
\begin{align}
p = \int_{1}^{1+r_0}dz \Psi_s(z) = 1 - e^{-Pe_z r_0} \approx Pe_z r_0\ .
\end{align}
The time $t_0$ to rotate about half of the angle $\theta_s$ is 
approximately $t_0 = \theta_s / Pe$. Therefore, we get
\begin{align}
  \label{mfpt:flowtime:two}
  T_F \approx \frac{4\pi}{N_r Pe Pe_z r_0}\ .
\end{align}
In this analysis, the convection time $T_F$ scales inversely with the
number of receptor patches $N_r$ and the P\'eclet number $Pe$.  As
$Pe$ increases, $T_F$ gets smaller than $T_D$ and then
dominates the overall outcome. Comparing \eq{mfpt:flowtime:two} to the
simulation data for $Pe \approx 0$ shows that this crossover occurs in
the range $Pe \approx 10^1 - 10^2$ and that the corresponding value of
$Pe$ increases with increasing receptor number $N_r$, exactly as
observed in the simulation data over the full range of $Pe$. However,
the exact scaling of this data is not $\sim 1/N_r$ for large $Pe$ as
predicted by \eq{mfpt:flowtime:two}. In practice, the decay is somehow
slower due to correlations between the height of the sphere at two
successive instances of a receptor pointing downwards, which we have
neglected in our analysis.
  
We briefly comment on the effect of the downward driving force, that
is $Pe_z$.  Above, we have found that in two cases, homogeneous
coverage from \eq{z-mfpt} and convection-dominated rotation from
\eq{mfpt:flowtime:two}, the MFPT scales inversely with $Pe_z$.  This
scaling behavior is indeed found in the simulations, except that for
very large values of $Pe_z$, the MFPT approximates a constant value
(data not shown).  The reason is that the larger $Pe_z$, the smaller
the mean time to fall below the height $z = 1 + r_0$. As indicated by
\eq{statsol}, then the sphere stays below this height until an
encounter occurs.  This implies that in this limit, the MFPT depends
only on rotational motion and the falling motion is irrelevant.

We now introduce spatially resolved ligands into the 2D-model.
\fig{fig:ldsetup}a shows the model definition: the ligand patches are
considered to have the same radius $r_d = r_p$ as the receptor patches and
they are located at a distance $d$ from each other. This results in a
one-dimensional ligand density given by $\rho_l = 2 r_d/d$.  The mean
first passage time will now also depend on the initial $x$-position,
$T = T(z_0, \theta, x | C)$, where $C$ is the hypersurface in
$(z,\theta, x)$ space where a receptor patch touches a ligand patch.
But similarly as in the above section in regard to initial
orientation, the dependence on the initial $x$-position is of minor
interest and therefore, we will discuss the MFPT
averaged over the initial position and orientation, denoted by
$\mean{T}_{\theta, x}$.

\fig{fig:ldsetup}b shows that by varying the P\'eclet number we can
identify the same three regimes for all ligand-densities as before.
For $Pe \rightarrow 0$ in the limit of pure diffusive transport,
$\mean{T}_{\theta, x}$ approaches a finite value, depending on
$\rho_r$ and $\rho_l$. With increasing $Pe$, $\mean{T}_{\theta,
  x}$ decreases monotonically and finally for $Pe \rightarrow
\infty$ reaches the value of the MFPT in the limit of homogeneous receptor and
ligand coverage. In contrast to above, however, in this limit the
shear flow not only restores rotational invariance of the sphere, but
in addition also translational invariance of the substrate.

\fig{fig:ligandresults}a provides more details for
$\mean{T}_{\theta, x}$ as a function of $\rho_l$ in the diffusive
limit ($Pe \approx 0$). We find that in the range $0.1 < \rho_l < 1$
the MFPT is almost not affected by ligand
concentration: as long as the ligand patches are sufficiently close to
each other, a receptor patch touching the wall will most probably find
a ligand before diffusing away again. The situation changes completely
with small ligand density.  For $\rho_l \ll 1$ the averaged mean first
passage time $\mean{T}_{\theta, x}$ scales with the ligand density
$\rho_l$ as $\mean{T}_{\theta, x} \propto 1/\rho_l^2 \propto d^2$.
This can be understood by calculating the position-averaged MFPT
$\mean{T}_x$ for a particle diffusing in an interval $[0,d]$ with
diffusion constant $D$, which gives $\mean{T}_x = d^2 / 12 D$.
This suggests that the quadratic scaling with $d$ results from the
diffusive motion between adjacent ligand patches.
\fig{fig:ligandresults}b summarizes our results for the dependence of
the 2D MFPT $\mean{T}_{\theta, x}$ on ligand
density $\rho_l$ and receptor density $\rho_r$ in the diffusive limit.
Clearly there exists a large plateau around the value for the case of
homogeneous coverage $\rho_r = \rho_l = 1$.  This implies that if
ligands and receptors patches are not too strongly diluted, the mean
encounter time is still close to the optimal value given by
\eq{z-mfpt}.  On the other hand if the number of receptor and/or
ligand patches is highly reduced the mean encounter time is strongly
increased.
 
\section{Movement in three dimensions}
\label{sec:3d}

We finally turn to the full 3D-situation, that is the sphere may
diffuse about all three axes as described by \eq{langevin-euler} and
\eq{rotation-update}. Receptors are located in spherical patches which
are randomly distributed over the sphere. Each receptor patch has a
radius $r_p$ and a height (capture length) $r_0$. That is the
appropriate generalization of the situation shown in \fig{fig:rpatch}
for the 2D-case.  Thus, for $N_r$ receptor patches the receptor density
is $\rho_r = 2\pi N_r(1 - \cos(r_p)) / 4\pi \approx N_r r_p^2 / 4$
(for $r_p \ll 1$).  In contrast to the preceeding sections where the
receptor patches could be regularly distributed over the circumference,
this is no longer possible on the surface of a sphere.
Therefore, we distribute the patches randomly over the sphere with
equal probability for each position, with a hard disk overlap
algorithm making sure that no two patches overlap \cite{hammer:92}.
One has to bear in mind that then for small $N_r$ two different
distributions may have slightly different binding properties. This
effect becomes weaker for larger $N_r$, therefore in the following we
will only use $N_r \geq 10$. The quantity we measure in our
simulations is now $\mean{T}_{\vec{\theta}}$ in the case of
homogeneous ligand coverage and $\mean{T}_{\vec{\theta},(x,y)}$ in the
case that the ligands are located in spherical patches on a
2D-lattice.  Thus, we average the MFPT over the initial orientations
and positions as explained above.

In order to explore the dependence of $\mean{T}_{\vec{\theta}}$ on
$N_r$ and $Pe$ we first simulated the receptor ligand encounter in the
case of homogeneous ligand coverage $\rho_l = 1$. In order to average
over the initial positions we started each run with a randomly chosen
initial orientation. After 100 runs we generated a new distribution,
thus averaging out also the effect of different receptor
distributions. In order to achieve reasonable statistics, we typically
used 100,000 runs.  Our results are shown in \fig{fig:3d}a. Again we
find three different regimes as a function of the P\'eclet number
$Pe$. This proves that qualitatively the basic results of the
2D-treatment remain valid in 3D. However, in detail there are
important differences. In contrast to the 2D results
presented above, $\mean{T}_{\vec{\theta}}$ in the limit $Pe
\rightarrow \infty$ is no longer given by \eq{z-mfpt} if $N_r$ is
small. That is due to the fact that for $Pe \rightarrow \infty$ the
receptor patches effectively behave as ring-like structures. The
rotation of such a ring about the $x$- or $y$-axis is not affected by
$Pe$ and thus still depends on diffusion.  For large $N_r$ the rings
cover the whole sphere and for $Pe \rightarrow \infty$
$\mean{T}_{\vec{\theta}}$ is again given by \eq{z-mfpt}.

In \fig{fig:3d}b we plot the $Pe \rightarrow 0$ limit of
$\mean{T}_{\vec{\theta}}$ as a function of the number of receptor
patches $N_r$, for different values of the capture radius $r_0$.  The
fitted straight line for $r_0 = 10^{-3}$ shows that
$\mean{T}_{\vec{\theta}}$ approximately behaves like
$\mean{T}_{\vec{\theta}} \propto 1/N_r$.  Neglecting effects of
curvature, the average distance between two receptors patches is $d
\propto (4\pi / N_r)^{1/2}$ and the mean time to diffuse that distance
is $t_{d} \propto {d}^2 \propto 1/N_r$.  This provides a simple
explanation for the observed scaling behavior. 
For high $N_r$, the
MFPT reaches a plateau value, given by \eq{z-mfpt}. This plateau value
depends on $r_0$ and is the smaller the larger $r_0$.  Also the
crossover from the asymptotic behavior at small $N_r$ to the plateau
at large $N_r$ is shifted with increasing capture radius $r_0$ towards
smaller $N_r$. 

In \fig{fig:3d}c we show the effect of a finite ligand density
$\rho_l$ at $Pe \approx 0$.  For the simulations we distributed the
ligands in circular patches of radius $r_d = 0.01$ on a quadratic
lattice with lattice constant $d$, thus, resulting in a ligand density
$\rho_l = \pi r_d^2 / d^2$. In our implementation, the intersection
between the receptor patch and the wall is approximated by an
appropriate circle, because it is easy to check if this circle
overlaps with the ligand patch. The fits given in \fig{fig:3d}c show
that for small $\rho_l$, the MFPT scales as
$\mean{T}_{\vec{\theta},(x,y)} \propto 1/\rho_l \propto d^2$.
Because the curves for different $N_r$ appear to be rather similar, in
the inset we plot the ratio of different pairs of these curves.  As
this results in approximately constant plateaus, we conclude that the
scaling with ligand density is hardly effected by $N_r$.  As in 2D,
the inverse scaling with ligand density can be understood in simple
terms by noting that the MFPT to diffusional capture scales like
$d^2$. At a coverage around $0.01$, saturation occurs as it did
for receptor coverage.

We finally discuss the influence of the receptor geometry described by
the parameters $r_0$ and $r_p$. Because $Pe$ changes the MFPT in a
monotonous way, it is sufficient to study the diffusive limit $Pe
\approx 0$. \fig{fig:3d:02}a and b show $\mean{T}_{\vec{\theta}}$
as a function of $r_p$ for $r_0 = 0.001$ and $r_0 = 0.01$,
respectively.  In order to obtain smooth curves, in this case only one
receptor distribution was used for all runs. We find that the curves
can be fitted well to the function
\begin{align}
  \label{receptor-fit}
  \mean{T(r_p)}_{\vec{\theta}} = \frac{a}{b + r_p} + T(z_0 =
  2| z_0 = 1 + r_0),
\end{align}
where the second term is the homogeneous result from \eq{z-mfpt}.
This means that even for vanishing receptor size $r_p \rightarrow 0$
the MFPT remains finite. This makes sense because
above we have shown that the effective patch
size is determined both by $r_p$ and $r_0$. In detail,
\fig{fig:rpatch} showed that capture occurs over the solid angle $2
\theta_0$ with $\theta_0(z) = \arccos(z/(1 + r_0))+ r_p$. For small
$r_0$ and $r_p$, this allows us to define an effective patch size
\begin{align}
  \label{def-reff}
r_p^{eff} =  \arccos(\mean{z}/(1+r_0)) + r_p \approx \arccos(1 - \rez{2} r_0) + r_p
\approx \sqrt{r_0} + r_p,
\end{align}
where we have used $\mean{z} = 1 + r_0 / 2$. Suppose now that the
sphere diffuses over the time $t_d$ until a receptor patch points
downwards, then it may encounter a ligand with a probability $p$ that
is given by the normalized area of one effective receptor patch:
\begin{align}
p = \rez{2}(1 - \cos(r_p^{eff})) \approx \rez{4} (\sqrt{r_0} + r_p)^2
\approx \rez{2}\sqrt{r_0} (\rez{2}\sqrt{r_0} + r_p).
\end{align}
If no encounter occurs, the sphere has to diffuse again a time $t_d$
until the next encounter can occur. This leads to the mean encounter
time $T = t_d/p$. Putting everything together gives \eq{receptor-fit}
with $a = 2t_d/(\sqrt{r_0})$ and $b = \rez{2}\sqrt{r_0}$. If checked
against our simulation results, we indeed find that the fit parameter
$b$ is an increasing function of $r_0$, but varies only slightly with
$N_r$.  The fit parameter $a$ scales approximately as $\sim 1/N_r$ and
varies with $r_0$, also consistent with the above analysis.  In
\fig{fig:3d:02}c $\mean{T}_{\vec{\theta}}$ is plotted as a
function of $r_p$ for several values of $r_0$ and $N_r = 30$. One
clearly sees that increasing $r_p$ has a much smaller impact on
$\mean{T}_{\vec{\theta}}$ than a comparable increase in $r_0$,
which is qualitatively well described by the preceeding analysis.

In \fig{fig:3d:02}a and b the receptor density is varied over almost
four orders of magnitude by changing $r_p$, but the largest measured
decrease for $\mean{T}_{\vec{\theta}}$ is only by a factor four.
In contrast, an increase of the receptor density by one order of
magnitude due to ten-fold more receptor patches leads to a decrease of
$\mean{T}_{\vec{\theta}}$ by almost also one order of magnitude.
However, this is only true as long as $N_r$ is not too large, as for
large $N_r$ $\mean{T}_{\vec{\theta}}$ saturates at the limiting
value of homogeneous receptor coverage (cf. \fig{fig:3d}b).  The
crossover from the $1/N_r$ behavior to the saturation should take
place when the average distance between two receptor patches
$d'\sim(4\pi/N_r)^{1/2}$ becomes comparable to the size of one
receptor patch. This corresponds to $r_p^{eff}\sim(4\pi/N_r)^{1/2}$ or
$N_r \sim 4\pi/(\sqrt{r_0} + r_p)$. This estimate predicts that the
crossover takes place between several tens to several hundreds of
receptor-patches, depending on $r_0$, in agreement with the data shown
in \fig{fig:3d}b.

\section{Summary and discussion}
\label{sec:discussion}

In this paper we have calculated the mean first passage times (MFPT) for
initial encounter between spatially resolved receptors on a Brownian
particle in linear shear flow and spatially resolved ligands on the
boundary wall. Our main results were obtained by repeated simulations
of the discretized Langevin equation \eq{langevin-euler}. Each data
point shown corresponds to at least 100,000 simulation runs. It is
important to note that these simulations are very time consuming
because we resolve objects of the size of $10^{-3} R$, that is for
$\mu$m-sized particles we resolve the nm-scale. 

In general, we found that the MFPT was always monotonically decreased
when the P\'eclet number was increased. That means that a particle
which is covered with receptors in a way that it binds well to ligands
already in the diffusive limit is even better suited to initiate
binding at finite shear rate. In our simulations we modeled the
receptor geometry using three parameters: the number of
receptor-patches $N_r$, the radius of the receptor patches $r_p$, and
the capture radius $r_0$.  The efficiency of binding is mainly
increased by $N_r$, but only up to a saturation value of the order of
hundred.  An increase of $r_p$ leads only to a weak enhancement of
binding efficiency.  The influence of $r_0$ to the MFPT is threefold:
i) it reduces the mean falling time, ii) it increases the effective
patch size, and iii) according to the stationary probability
distribution for the $z$-direction, it becomes more probable for the
sphere to be within the encounter zone when $r_0$ is increasing.  An
additional but more indirect effect of receptor protrusions is that
the further the cell is away from the wall, the faster it can rotate
(even in the diffusive limit) due to the larger mobility.  As shown by
\eq{mfpt:add} rotations play a role only within binding range, i.\,e.,
for $z < 1 + r_0$. Therefore, a large $r_0$ lets the cell also benefit
from faster rotations.  Summarizing our findings in regard to receptor
geometry we conclude that the most efficient design for particle
capture under flow is to cover the particle with hundreds of
receptor patches ($N_r$ above threshold), each with a rather small
area (small $r_p$), but formed as a protrusion (large $r_0$).

Indeed, this strategy seems to be used by white blood cells, which have
evolved intriguing mechanisms both on the molecular and cellular scale
in order to adhere effectively to the endothelium under the conditions
of hydrodynamic flow. The typical size of white blood cells is $R
\approx 5\ \mu$m and they are covered with a few hundreds of
protrusions (\textit{microvilli}) with the receptors (most notably
L-selectin) localized to the microvilli tips \cite{c:chen99}. In
general, the microvilli of white blood cells are much more complex
than the parameter $r_0$ in our model: they are rather long (typical
length 350 nm, that is $R / 15$) and have their own physical
properties (e.g., very flexible in the transverse direction and
viscous in the longitudinal direction) \cite{c:shao98}.  Nevertheless,
it is striking that elevation of the receptors above the main cell
surface seems to be a major design principle for white blood cells.
In fact, the same strategy appears to be used also by malaria-infected
red blood cells, which are known to develop a dense coverage with
elevated receptor patches (\textit{knobs}) on the cell surface
\cite{c:bann03,c:naga00,c:amin05}. A typical value for the cell radius
is 3.5 $\mu$m \cite{c:sure05}. The knobs have a typical height of 20
nm, a radius of about 90 nm and a distance of 200 nm (for red blood
cells infected by single parasites) \cite{c:naga00}. This dense and
elevated coverage suggests that like the white blood cells, the
malaria-infected red blood cells also function in the regime of
homogeneous coverage.

In order to discuss the motion of white blood cells in more detail, it
is instructive to consider the parameters for a typical flow chamber
experiment. In aqueous solution and at room temperature, $\rho =$
g/cm$^3$, $\eta = 10^{-3}$ Pa s, and $T = 293 K$. Then, the
dimensionless parameters determining cell motion become
\begin{align}
  \label{peclets}
  Pe = 4.67 R^3 \dot\gamma, \;\;\;
  f = 2.17 \frac{R \Delta \rho}{\dot\gamma}, \;\;\; 
  Pe_z = 10.16 R^4 \Delta \rho, \;\;\;
  \Delta t = \frac{Pe}{\dot\gamma} = 4.67 R^3 s, 
\end{align}
where $R$ is given in $\mu m$, $\Delta \rho$ in units of g/cm$^3$ and
the shear rate $\dot{\gamma}$ in units of 1 per seconds; $\Delta t$ is
the diffusive time scale.  For leukocytes in flow chambers we
typically have $R = 5$, $\dot \gamma = 100$ and $\Delta \rho = 0.05$,
thus, for the two P\'eclet numbers we get $Pe = 6 \times 10^4$ and
$Pe_z = 317$, respectively. Then, $f = Pe_z / Pe = 0.005$, that is the
effect of hydrodynamic deterministic motion will be very strong.  The
experimental time scale is given by the time for transversing the
field of view, which is about 3 s at a shear rate of 100 Hz and length
of 670 $\mu$m. The diffusive time scale $\Delta t$ for leukocytes is
about 600 s (10 min), which reflects their large size and shows that
diffusive motion is by far not sufficient to initiate binding.
Binding becomes more favorable in the presence of convection.  For a
start height of one radius above the wall ($z_0 = 2$), our
calculations give a MFPT of about 5 s, that is much less than the
diffusive time. However, this is still much larger than the
experimental time scale.  This proves that only those cells have a
chance to bind that flow very close to the wall, exactly as observed
experimentally. In vivo, white blood cells therefore depend also on
other mechanisms driving them onto the substrate, including contact
and hydrodynamic interactions with other cells. These effects have
been studied in detail before.  For example, Munn and coworkers have
shown that adhesion of leukocytes close the the vessel wall in
post-capillary venules is enhanced by red blood cells passing them
\cite{munn:03}. King and Hammer have shown, using an algorithm capable
of simulating several cells, that already adherent leukocytes can
recruit other leukocytes via hydrodynamic interactions
\cite{hammer:01a}.  The results presented here, when specified to
leukocytes, show that indeed these mechanisms are crucial for
effective leukocyte capture under flow.

Our results also suggest that leukocytes are sufficiently large that
thermal fluctuations are not dominant. This changes when studying
smaller particles, e.g., receptor-covered spheres with $R \approx 1\
\mu$m, whose binding also has been investigated with flow chambers
\cite{pierres:98,pierres:01}. \eq{peclets} shows that the P\'eclet
numbers scale strongly with particle radius $R$, therefore, these beads
are subject to much stronger thermal fluctuations than leukocytes.  In
Ref.~\cite{pierres:01} it has been verified that indeed in equilibrium
such particles obey the barometric distribution from \eq{statsol}. In
Ref.~\cite{pierres:98} it was found that the adhesion probability
$p_{ad}$ is proportional to the ligand-density, $p_{ad} \sim \rho_l$.
With $p_{ad} \sim 1/T$ it follows that $T \sim 1/\rho_l$ as found by
our simulations in the limit of low ligand densities.

Throughout this paper we have considered the generic case of a
constant downward acting force due to a density difference between the
sphere and the surrounding fluid. In future work it might be
interesting to examine also other forces which can easily be done in
the framework presented here.  As the addition formula \eq{mfpt:add}
for falling and rotational MFPT was not derived under the assumption
of a specific force, it is also true for non-constant forces. For
general potential forces the falling time \eq{z-mfpt} has then to be
replaced by \eq{mfpt-potential}. Also the rotational MFPT is
influenced by a vertical force via the stationary height distribution.
Neglecting gravitational force and considering only short-ranged forces
like van der Waals or electrostatic forces would result in infinite
MFPTs for the setup of the halfspace.  This problem, however, can be
solved by using an additional wall acting as an upper boundary
\cite{jones:04}.

In this paper we assumed a rigid Brownian particle. For cells, elastic
deformations might be relevant. For free flow, a simple scaling
estimate shows that the critical value for the shear rate leading to
substantial deviations from the spherical shape is $(E h)/(\eta R)$
\cite{schwarz:00}, where $E = 100$ Pa and $h = 100$ nm are Young
modulus and thickness of the cellular envelope, respectively. The fact
that the Young modulus E appears here indicates that cells tend to
passively deform less than vesicles, whose elasticity is characterized
rather by the bending rigidity \cite{c:seif96,c:suku01}. The scaling
estimate leads to a critical shear rate of $10^3$ Hz, which is above
the value of a few $10^2$ Hz (corresponding to $Pe \approx 10^5$ for
white blood cells) which often provides an upper limit in flow chamber
experiments.  Similar but more complicated scaling arguments can be
made for lubrication forces which arise when the cell approaches the
wall \cite{sekimoto:93}.

To fully understand the rate of association between a receptor-covered
particle in shear flow and a ligand-covered wall, our analysis should
be completed by the implementation of an adhesion scenario, which in
general should also include molecular determinants like residence
times and receptor flexibility. If one assumes that a bond between two
encountering molecules is formed with a certain rate, then, the MFPT
for encounter as reported here should be a good approximation for the
mean adhesion time in the limit of zero shear rate, because in this
limit the duration of each encounter should be sufficiently long for
the formation of an adhesion contact. Then, the proper knowledge of the
MFPT could also be used to design a cell sorting experiment. Suppose
one has a mixture of different cells each bearing some receptors and
the wall is covered with one kind of ligand. Then, the cells are flowed
into the chamber and flow is stopped. Certainly, only cells that bear
receptors which fit to the ligands can attach to the wall. If the flow
is then turned on again, the attached cells will be separated from the
other cells. If the no-flow period is much shorter than the MFPT, only
a few cells can attach. If the no-flow period is much longer than the
MFPT, attached cells might already start to spread and are therefore
difficult to remove. Only if the no-flow period is of the order of
MFPT one gets an appreciable number of weakly attached cells. In this
sense our theoretical analysis might be essential for appropriate
biotechnological applications.

\begin{acknowledgments}
  We thank Reinhard Lipowsky for general support.  This work was
  supported by the German Research Foundation (DFG) through the Emmy
  Noether Program and by the Center for Modelling and Simulation in
  the Biosciences (BIOMS) at Heidelberg.
\end{acknowledgments}

\appendix

\section{Implementation of friction and mobility matrices}
\label{appendix:mobility}

For the numerical implementation of the friction and mobility tensors
for a sphere in linear shear flow above a wall we use the results from
Refs.~\cite{jones:92,jones:98}. This implementation procedure has been
described and tested in detail in Ref.~\cite{jones:98}. In this
appendix, we briefly summarize it for the sake of completeness.
  
Writing the friction tensors in terms of irreducible tensors formed
from $\delta_{ij}, \epsilon_{ijk}, \boldv{k}$ defines the scalar
friction functions.  In the case that the normal vector to the wall is
$\boldv{k} = \boldv{e}_z$, these tensors read
\label{frictionfunctions}
\begin{align*}
  \zeta^{tt} &= \left(\begin{array}{ccc}
    \psi^{tt}&0&0\\
    0&\psi^{tt}&0\\
    0&0&\phi^{tt}
  \end{array}\right), \;\;\;
  \zeta^{tr} = \psi^{tr} \left(\begin{array}{ccc}
    0&1&0\\
    -1&0&0\\
    0&0&0
  \end{array}\right)=-{\zeta^{rt}}^T,\;\;\;
  \zeta^{rr} = \left(\begin{array}{ccc}
    \psi^{rr}&0&0\\
    0&\psi^{rr}&0\\
    0&0&\phi^{rr}
  \end{array}\right),\\
  \zeta^{td}_\alpha &= \left(\begin{array}{ccc}
    -\rez{3}\delta_{\alpha 3}\phi^{td}&0&\rez{2}\delta_{\alpha 1}\psi^{td}\\
    0&-\rez{3}\delta_{\alpha 3}\phi^{td}&\rez{2}\delta_{\alpha 2}\psi^{td}\\
    \rez{2}\delta_{\alpha 1}\psi^{td}&\rez{2}\delta_{\alpha 2}\psi^{td}&
    \frac{2}{3}\delta_{\alpha 3}\phi^{td}
  \end{array}\right),\ha
  \zeta^{rd}_\alpha = \rez{2}\psi^{rd}\left(\begin{array}{ccc}
    0&0&\epsilon_{3 \alpha 1}\\
    0&0&\epsilon_{3\alpha 2}\\
    \epsilon_{3 \alpha 1}&\epsilon_{3 \alpha 2}&0
  \end{array}\right),\\
  \zeta^{dt}_\alpha &= \left(\begin{array}{ccc}
    \rez{2}\delta_{\alpha 3}\psi^{dt}&0& -\rez{3}\delta_{\alpha 1}\phi^{dt}\\
    0&\rez{2}\delta_{\alpha 3}\psi^{dt}& -\rez{3}\delta_{\alpha 2}\phi^{dt}\\
   \rez{2} \delta_{\alpha 1}\psi^{dt}&\rez{2}\delta_{\alpha 2}\psi^{dt}& 
   -\frac{2}{3}\delta_{\alpha 3}\phi^{dt}
  \end{array}\right),\ha
  \zeta^{dr}_\alpha = \rez{2}\psi^{dr}\left(\begin{array}{ccc}
    0&\delta_{\alpha 3}&0\\
    -\delta_{\alpha 3}&0&0\\
    -\delta_{\alpha 2}&\delta_{\alpha 1}&0
  \end{array}\right).
\end{align*}
This defines the scalar friction functions $\phi^{tt}, \psi^{tt},
\psi^{tr}, \psi^{rt}, \phi^{rr}, \psi^{rr}, \phi^{td}, \psi^{td},
\psi^{dr}$. The scalar friction functions $\phi$ depend only on the
inverse distance of the sphere from the wall, that is the
dimensionless variable $t = R/z$, which takes values from the interval
$[0,1]$.  The friction functions can be expanded in powers of $t$.
The numerically obtained first 20 coefficients of such a series
expansion of the dimensionless scalar friction functions
\label{frictionfunctions2}
\begin{align*}
  \hat\phi^{tt} &= \phi^{tt}/6\pi\eta R,\ha
  \hat\psi^{tt} = \psi^{tt}/6\pi\eta R,\ha
  \hat\phi^{rr} = \phi^{rr}/8\pi\eta R^3,\ha\\
  \hat\psi^{rr} &= \psi^{rr}/8\pi\eta R^3,\ha
  \hat\psi^{tr} = \psi^{tr}/8\pi\eta R^2 = -\hat\psi^{rt}
\end{align*}
are tabulated in Ref.~\cite{jones:92}. 
For the other three dimensionless scalar friction functions 
\begin{align*}
  \hat\phi^{dt} =\phi^{dt}/6\pi\eta R^2 = \hat\phi^{td},
  \ha\hat\psi^{dt} =\psi^{dt}/6\pi\eta R^2 = \hat\psi^{td},
  \ha\hat\psi^{dr} =\psi^{dr}/8\pi\eta R^3 = -\hat\psi^{rd} 
\end{align*}
the first 32 coefficients of a series expansion in powers of $t$ are
tabulated in Ref.~\cite{jones:98}.  For small values of $t$ the series
expansion converges quite well and only a few coefficients are needed
to obtain accurate results. However, for $t\rightarrow 1$,
i.\,e., close to the wall, the friction functions are better described
in a lubrication expansion, which reads
\begin{align*}
  \hat\phi \approx C_1 \frac{t}{1-t} + C_2\ln(1-t) + C_3 + C_4\frac{1-t}{t}\ln(1-t)+
  \mathcal{O}(1-t).
\end{align*}
The coefficients $C_1,C_2,C_3,C_4$ for the eight friction functions
defined above can be found in Ref.~\cite{jones:98}.
%are listed in the following table
%\begin{center}
%\begin{tabular}{c||c|c|c|c|c|c|c|c|}
%  &$\hat \phi^{tt}$&$\hat \psi^{tt}$&$\hat \phi^{rr}$
%  &$\hat \psi^{rr}$&$\hat \psi^{tr}$&$\hat \phi^{dt}$
%  &$\hat \psi^{dt}$&$\hat \psi^{dr}$\\
%  \hline
%  $C_1$&$1$&$0$&$0$&$0$&$0$&$-1$&$0$&$0$\\
%  \hline
%  $C_2$&$-\frac{1}{5}$&$-\frac{8}{15}$&$0$&$-\frac{2}{5}$
%  &$\frac{1}{10}$&$-\frac{4}{5}$&$\frac{14}{15}$&$\frac{1}{5}$\\
%  \hline
%  $C_3$&$0.97127$&$0.95429$&$\zeta(3)$&$0.37089$&$0.19295$&$-0.30697$&$1.23538$&$0.18719$\\
%  \hline
%  $C_3$&$-\frac{1}{21}$&$-\frac{64}{375}$&$0$&$-\frac{66}{125}$&$\frac{43}{250}$
%  &$-\frac{13}{21}$&$\frac{442}{375}$&$-\frac{2}{125}$\\
%  \hline
%\end{tabular}
%\end{center}
In order to match the two limit cases, the the asymptotic expansion of
the $t\rightarrow 1$ limit is subtracted from the friction functions
\begin{align*}
  \hat\phi(t) = \sum\limits_{n = 0}^\infty f_n t^n,
\end{align*}
leading to a new series expansion:
\begin{align*}
  \hat\phi(t) &- C_1 \frac{t}{1-t} - C_2\ln(1-t) - C_4\frac{1-t}{t}\ln(1-t) \\
  &= f_0 + C_4 + \sum\limits_{n = 1}^\infty
  \left(f_n - C_1 + \frac{C_2}{n} - \frac{C_4}{n(n+1)}\right)t^n 
  =: \sum\limits_{n = 0}^\infty g_n t^n.
\end{align*}
This series is truncated at $n_{max} = N$ and the coefficients $g_n$
are calculated from the coefficients $f_n, C_i$.  Next the
coefficients $g_n$ ($n = 0,\ldots, N$) are not used to calculate the
Taylor sum, but rather to calculate the Pad\'e approximant to this
function.  The Pad\'e approximant is given as
\begin{align*}  
  \mathcal{P}_N(t) = \frac{a_0 + a_1t + a_2t^2 + \ldots + a_Nt^N}{1 + b_1t + b_2t^2 
    + \ldots + b_Nt^N}
\end{align*}
where the coefficients $a_i, b_j$ are the solution to 
\begin{align*}
  \sum\limits_{n=1}^Nb_ng_{N-n+k} = -g_{n+k},\ha \sum\limits_{n=1}^kb_ng_{k-n} = a_k,\ha
  k = 1,\ldots,N.
\end{align*}
Finally the numerically implemented friction functions become
\begin{align}
  \label{frictnumeric}
    \hat\phi(t) = C_1 \frac{t}{1-t} + C_2\ln(1-t) + C_4\frac{1-t}{t}\ln(1-t)
    + \mathcal{P}_N(t).
\end{align}
For the calculation of the coefficients $a_i, b_j$ of the Pad\'e
approximant we use the algorithm provided by the Numerical Recipes
\cite{numerical:c}.

Having implemented the scalar friction functions, the implementation
of the mobility tensors proceeds by substituting $\zeta
\leftrightarrow \mu, \phi \leftrightarrow \alpha,\psi \leftrightarrow
\beta$ in the above decomposition of the friction tensors. This
defines the scalar mobility functions $\alpha^{tt}, \beta^{tt},
\alpha^{rr}, \beta^{rr}, \beta^{tr}, \alpha^{dt}, \beta^{dt},
\beta^{dr}$.  Using \eq{mobilitymatrix} the dimensionless scalar
mobility functions can be calculated from the scalar friction
functions:
\label{mobilityfunctions2}
\begin{align*}
  \hat\alpha^{tt} &= 1/\hat\phi^{tt}, \ha
  \hat\beta^{tt} = 
  \frac{\hat\psi^{rr}}{\hat\psi^{tt}\hat\psi^{rr} - \frac{4}{3}(\hat\psi^{tr})^2}\\
  \hat\alpha^{rr} &= 1/\hat\phi^{rr}, \ha
  \hat\beta^{rr} = 
  \frac{\hat\psi^{tt}}{\hat\psi^{tt}\hat\psi^{rr} - \frac{4}{3}(\hat\psi^{tr})^2}\\
  \hat\beta^{tr} &= 
  -\frac{4}{3}
  \frac{\hat\psi^{tr}}{\hat\psi^{tt}\hat\psi^{rr} - \frac{4}{3}(\hat\psi^{tr})^2}\\
  \hat\alpha^{dt} &= -\hat\phi^{dt} \hat\alpha^{tt},\ha 
  \hat\beta^{dt} = -\hat\psi^{dt}\hat\beta^{tt}-\hat\psi^{dr}\hat\beta^{tr},\ha 
  \hat\beta^{dr} = -\frac{3}{4}\hat\psi^{dt}\hat\beta^{tr}-\hat\psi^{dr}\hat\beta^{rr}.
\end{align*}
In \fig{abbmobility} we use our implementation to plot the eight
dimensionless mobility functions.

The limit of an unbounded flow corresponds to $t \rightarrow 0$ and results in
\begin{align}
  \label{friction-nowall}
  \zeta^{tt} = 6\pi\eta R I ,\;\;
  \zeta^{rr} = 8\pi\eta R^3 I ,\;\;
  \zeta^{tr} = \zeta^{rt} = \zeta^{rd} = \zeta^{td} = 0
\end{align}
where $I$ is the unity matrix. Thus eq. (\ref{frictionmatrix2}) reduces to
\begin{align}
  \boldv{F}^H = 6\pi\eta R \left(\boldv{U} - \boldv{U}^\infty\right),\ha
  \boldv{T}^H = 8\pi\eta R^3\left(\boldv{\Omega} - \boldv{\Omega}^\infty\right).
\end{align}
which are the well-known Stokes laws for the friction force and torque
exerted on a sphere moving in a fluid with relative velocity
$\boldv{U} - \boldv{U}^\infty$. For the linear shear flow considered
here, $\boldv{U}^\infty = \dot\gamma z \boldv{e}_x$ and
$\boldv{\Omega}^\infty = \dot\gamma \boldv{e}_y / 2$.

\section{Relation to the Smoluchowski equation}
\label{appendix:gradient_term}

The probability distribution $\Psi(\boldv{X},t)$ of a Brownian
particle subject to external force/torque $\boldv{F}$ satisfies a
continuity equation $\partial_t \Psi + \nabla \cdot \boldv{J} = 0$.
The probability flux $\boldv{J}$ contains a diffusive and a convective
part \cite{vankampen:92}:
\begin{align}
  \label{flux}
  J_i = - D_{ij} \partial_j \Psi + M_{ij} F_j \Psi 
\end{align}
where $\mathsf{D}$ and $\mathsf{M}$ are diffusion and mobility
matrices, respectively, and $\boldv{F}$ is external force. In
equilibrium, the flux has to vanish and the probability distribution
has to become the Boltzmann distribution. This leads to the Einstein
relation $\mathsf{D} = k_B T \mathsf{M}$, which is a special case of
the fluctuation-dissipation theorem. Using \eq{flux} and the
Einstein relation in the continuity equation leads to the 
Smoluchowski equation \cite{edwards}:
\begin{align}
  \label{smoluchowski}
  \partial_t \Psi = \partial_i \left(
  \mathsf{M}_{ij}(k_BT \partial_j \Psi - \boldv{F}_j \Psi )\right).
\end{align}
We now will derive the equivalent Langevin equation. In the case of
constant mobility (\textit{additive noise}), e.\,g.,  $\mathsf{M}_{ij}
= \delta_{ij}$, the appropriate Langevin equation is given by
\begin{align}
  \label{langevin-constant}
  \partial_t \boldv{X}_t = \mathsf{M}\boldv{F} + \boldv{g}_t^S,
\end{align}
where $\boldv{g}_t^S$ is a Gaussian white noise term and the
Stratonovich interpretation is used as explained in the main text.
However if $\mathsf{M}$ depends on $\boldv{X}$ (\textit{multiplicative
noise}), an additional drift term occurs in the Langevin equation
\begin{align}
  \label{langevin-drift}
  \partial_t \boldv{X}_t = \mathsf{M}\boldv{F} + k_BT \boldv{Y} + \boldv{g}_t^S.
\end{align}
The following derivation of the drift term $\boldv{Y}$ proceeds in two
steps \cite{edwards}. First we perform a coordinate transformation
which makes the noise additive.  In the case of additive noise the
Langevin equation (\ref{langevin-constant}) and the Fokker-Planck
equation (\ref{smoluchowski}) are equivalent. Then starting from the
Fokker-Planck equation in the new coordinates we perform the
transformation back to the old coordinates. Requiring the transformed
Fokker-Planck equation to be of the same form as in \eq{smoluchowski},
determines the drift term $\boldv{Y}$.

As we use the Stratonovich interpretation for the noise process the
usual rules for differentiation and integration apply and we can
perform the following coordinate transformation
\begin{align}
  \label{trafo-01}
  \boldv{X}' = \int\limits^{\boldv{X}(t)} \mathsf{S}(\boldv{X}'') \de 
  \boldv{X}'',
\end{align}
with some regular matrix $\mathsf{S}$.
The Langevin equation for the transformed coordinates then reads
\begin{align}
  \label{langevin-02}
  \partial_t \boldv{X}_t' = \mathsf{S} \partial_t \boldv{X}_t 
  =  \mathsf{SM}\boldv{F} + k_BT\mathsf{S}\boldv{Y} + \mathsf{S}\boldv{g}_t^S. 
\end{align}
From the requirement that $\mathsf{M'}_{ij} = \delta_{ij}$, that is
\begin{align}
  \label{mean}
  \mean{\mathsf{S}\boldv{g}_t\mathsf{S}\boldv{g}_t} \stackrel{!}{=} 
  2 k_BT \mathsf{E}, \ha \mathsf{E}_{ij} := \delta_{ij},
\end{align}
we can fix $\mathsf{S}$ to be the inverse of a matrix $\mathsf{B}$ with
\begin{align}
  \label{def-B}
  \mathsf{S} = \mathsf{B}^{-1},\ha
  \mathsf{M} = \mathsf{BB}^T \ha \Leftrightarrow
  \ha  \mathsf{M}_{ij} = \mathsf{B}_{ik}\mathsf{B}_{jk}.
\end{align}
As $\mathsf{M}$ is a symmetric positive definite matrix, it is always
possible to find a matrix $\mathsf{B}$ with $\mathsf{M} =
\mathsf{BB}^T$. Defining
\begin{align}
  \label{def-fnew}
  \boldv{F}' := \mathsf{B}^T\boldv{F} + k_BT\mathsf{S}\boldv{Y}, \ha   
  \tilde{\boldv{g}}_t^S := \mathsf{S}\boldv{g}_t^S = \mathsf{B}^{-1}\boldv{g}_t^S,
\end{align}
the new Langevin equation for the primed coordinates and with additive noise reads 
\begin{align}
  \label{langevin-02a}
  \partial_t \boldv{X}_t' = \mathsf{M}'\boldv{F}' + \tilde{\boldv{g}}_t^S. 
\end{align}
The corresponding probability distribution $\Psi'(\boldv{X}', t)$ is
the solution of the Smoluchowski equation
\begin{align}
  \label{smol-02}
  \partial_t \Psi'(\boldv{X}', t) = \partial_k' \delta_{ki} 
  (k_BT \partial_i' \Psi' -F'_i\Psi').
\end{align}
Next we transform (\ref{smol-02}) back to the unprimed coordinates.
The preservation of probability requires that
\begin{align}
  \label{proppreserv}
  {\Psi'}(\boldv{X}', t) = J\Psi(\boldv{X}, t)
\end{align}
where $J$ is the Jacobian of the coordinate transformation \cite{lax:66}: 
\begin{align}
  \label{trafo-02}
  J := \mathrm{det}\left(\frac{\partial X_i}{\partial X_j'}\right) 
     = \mathrm{det}\left(\mathsf{B}\right), \ha
     \frac{\partial X_i}{\partial X_j'} &= \mathsf{B}_{ij}.
\end{align}
Inserting (\ref{proppreserv}) into (\ref{smol-02}) gives
\begin{align}
  \label{smol-mid}
  \partial_t \Psi' = J\partial_t \Psi = 
  \partial_k'(k_BT\partial_k' \Psi' - F_k' \Psi') = 
  k_BT\partial_k'\partial_k' J\Psi - \partial_k' F_k' J \Psi.
\end{align}
Dividing by $J$ we obtain for the first term on the right hand side of (\ref{smol-mid})
\begin{align*}
  J^{-1}\partial_k'\partial_k' J\Psi &= 
  J^{-1}(\partial_k'\partial_k' J)\Psi + 2J^{-1}(\partial_k'J)\partial_k'\Psi
  + \partial_k'\partial_k'\Psi\\
%  &= (\partial_j \mathsf{B}_{jk})(\partial_l \mathsf{B}_{lk})\Psi + 
%  \mathsf{B}_{jk}(\partial_j\partial_l \mathsf{B}_{lk})\Psi
%  + 2(\partial_j\mathsf{B}_{jk})\mathsf{B}_lk\partial_l \Psi + 
%  \mathsf{B}_{jk}\partial_j(\mathsf{B}_{lk}\partial_l \Psi)\\
%  & =  (\partial_j\mathsf{B}_{jk}(\partial_l \mathsf{B}_{lk}))\Psi + 
%  \partial_j(\mathsf{B}_{jk}\mathsf{B}_{lk}\partial_l\Psi)
%  + \mathsf{B}_{jk}(\partial_l\mathsf{B}_{lk}) \partial_j\Psi\\
  &= \partial_j(\mathsf{B}_{jk}\mathsf{B}_{lk}\partial_l\Psi + 
  \mathsf{B}_{jk}(\partial_l\mathsf{B}_{lk})\Psi).
\end{align*}
Here we made use of the identities
\begin{align}
  \label{trafo-03}
  J^{-1}\grad' J &= \grad \mathsf{B}^T, J^{-1}\partial_i' J = 
  \partial_j \mathsf{B}_{ji},\ha 
  \grad ' = \mathsf{B}^T\grad,\\
  \nonumber
  J^{-1}\partial_i'\partial_j' J &=J^{-1}\partial_i'(JJ^{-1})\partial_j'J = 
  J^{-1}(\partial_i'J)J^{-1}\partial_j'J + 
  \partial_i'(J^{-1}\partial_j'J) \\
  \nonumber
  &= (\partial_k \mathsf{B}_{ki})\partial_l \mathsf{B}_{lj} + 
  \mathsf{B}_{li}\partial_l\partial_k \mathsf{B}_{kj}.
\end{align}
Again using the identity (\ref{trafo-03}) the second term of the right hand side of 
(\ref{smol-mid}) can be evaluated to be
\begin{align*}
  J^{-1}\partial_k'  F_k' J \Psi &= J^{-1}(\partial_k'J) F_k' \Psi + 
  \partial_k' F_k' \Psi%\\
  %& = (\partial_j\mathsf{B}_{jk}) F_k' \Psi + \mathsf{B}_{jk}\partial_j(F_k' \Psi) 
  = \partial_j (\mathsf{B}_{jk} F_k' \Psi).
\end{align*}
Adding both terms and inserting the definitions (\ref{def-B}) and (\ref{def-fnew}) we have
\begin{align*}
  \partial_t \Psi = \partial_j\left(k_BT\mathsf{M}_{jl}\partial_l\Psi + 
  k_BT\mathsf{B}_{jk}(\partial_l\mathsf{B}_{lk})\Psi
  - \mathsf{M}_{jl}F_l - k_BT Y_j  \Psi \right).
\end{align*}
Comparing this with the required result (\ref{smoluchowski}) we can
read off $\boldv{Y}$
\begin{align*}
  \boldv{Y} = \mathsf{B}\grad\mathsf{B}^T,\ha Y_i = 
  \mathsf{B}_{ik}(\partial_l\mathsf{B}_{lk}).
\end{align*}
Finally shifting $\partial_t\boldv{X}_t \rightarrow
\partial_t\boldv{X}_t - \boldv{U}^\infty$ we obtain the Langevin
equation as given by \eq{langevin-01} combined with \eq{matrixB}.

\section{Euler algorithm for a sphere above a wall}
\label{appendix:euler}

In order to solve \eq{langevin-04} numerically we use an Euler
algorithm.  As the physical situation
requires to use the Stratonovich interpretation of the noise term
$\boldv{g}_t^S$, the displacement $\Delta \boldv{X}$ of a particle
from time $t$ to time $t+\Delta t$ depends on the position of the
particle at time $t + (1/2)\Delta t$, which is not known at time $t$.
As usual, this problem is solved by rewriting the Langevin equation in
the It\^o-version.  Then the noise term can be evaluated at time $t$
and as a compensation an additional drift term $\partial_l
(\mathsf{B}_{ik})\mathsf{B}_{lk}$ is added to \eq{langevin-04}
\cite{vankampen:92}. Because $\mathsf{B}_{kl}^T\partial_l
(\mathsf{B}_{ik}) + \mathsf{B}_{ik}\partial_l (\mathsf{B}^T_{kl}) =
\partial_l (\mathsf{B}_{ik}\mathsf{B}_{kl}^T) = \partial_l
\mathsf{M}_{il}$, we arrive at \eq{langevin-ito}. In this equation,
the random displacements $\boldv{g(\Delta t)}$ must satisfy
\begin{align}
  \label{rand-displ}
  \mean{\boldv{g}(\Delta t)} = 0, \ha 
  \mean{\boldv{g}(\Delta t)\boldv{g}(\Delta t)} = 2 \mathsf{M} \Delta t.
\end{align}
Following Ref.~\cite{ermak:78}, $\boldv{g}_i(\Delta t)$ is calculated
from a weighted sum of normal deviate random numbers $\bar x_i
\rightarrow \{x_i\}$ satisfying $\mean{x_i} = 0, \mean{x_ix_j} =
2\delta_{ij}\Delta t$. This sum is given by
\begin{align*}
  \boldv{g}_i(\Delta t) = \sum\limits_{j = 1}^i \mathsf{B}_{ij} \bar x_j
\end{align*}
where the weighting factors are the elements of the matrix
$\mathsf{B}$ defined in (\ref{def-B}).  They can recursively be
calculated according to
\begin{align*}
  \mathsf{B}_{ii} =\left(\mathsf{M}_{ii} -  \sum\limits_{k =
  1}^{i-1}\mathsf{B}_{ik}^2\right)^{\frac{1}{2}},\ha  \mathsf{B}_{ij}
  = \left(\mathsf{M}_{ij} - \sum\limits_{k =
  1}^{j-1}\mathsf{B}_{ik}\mathsf{B}_{jk}\right)/\mathsf{B}_{jj}, i >
  j, \ha \mathsf{B}_{ij} = 0, i < j.
\end{align*}
In the case of a sphere above a wall we obtain the following
dimensionless weighting factors (cf. \cite{jones:92a})
\begin{align}
  \label{diffsqrt}
  \mathsf{\hat B}_{11} &= \sqrt{\hat\beta^{tt}},\ha
  \mathsf{\hat B}_{22} = \sqrt{\hat\beta^{tt}},\ha
  \mathsf{\hat B}_{33} = \sqrt{\hat\alpha^{tt}},\ha
  \mathsf{\hat B}_{42} = -\mathsf{\hat B}_{51} 
  = -\frac{3}{4}\frac{\hat\beta^{tr}}{\sqrt{\hat\beta^{tt}}},\\
  \mathsf{\hat B}_{44} &= \mathsf{\hat B}_{55} 
  = \frac{3}{4}\rez{\sqrt{\hat\beta^{tt}}}  
  \left(\frac{4}{3}\hat\beta^{tt}\hat\beta^{rr} 
  - (\hat\beta^{tr})^2\right)^{\frac{1}{2}}
  \equiv \sqrt{\frac{3}{4\hat\psi^{rr}}},\ha
  \mathsf{\hat B}_{66} = \rez{2}\sqrt{3\hat\alpha^{rr}}. 
\end{align}
As pointed out in Ref.~\cite{honerkamp:94}, using the Euler method,
instead of normal deviate random variables any uncorrelated random
variable $\bar x_i \rightarrow \{x_i, i = 1,\ldots,6\}$ can be chosen,
as long as they fulfill the required relation for the first moments
$\mean{x_i} = 0, \mean{x_ix_j} = 2\delta_{ij}\Delta t$.  Thus, it is
much faster to generate the random numbers according to $\bar x_i =
\sqrt{12 \Delta t}(\xi_i - 0.5)$, with $\xi_i, i = 1,\ldots,6$ being
uncorrelated random variables uniformly distributed in $[0,1]$.  For
the calculation of the random numbers we use the pseudo random number
generator {\sf ran3} from the Numerical Recipes \cite{numerical:c}.

Calculating the new configuration after each time-step using
(\ref{langevin-euler}) is straightforward for the spatial degrees of
freedom. For the update of the orientation of the sphere we use a
coordinate system spanned by three orthonormal basis-vectors
$\{\vec{n}_i|i=1,2,3; (\vec{n}_i)_j = \delta_{ij}\}$. The origin of
this coordinate system shall be identical with the center of mass of
the sphere and the relative orientation of this system and of the
sphere are kept fixed.  Given then an orientation update form
(\ref{langevin-euler}) $\vec{\theta} :=
(\Delta\boldv{X}_4,\Delta\boldv{X}_5,\Delta\boldv{X}_6)$, we decompose
each of the basis vectors $\vec{n}_i$ into a component parallel to
$\vec{\theta}$ denoted by $\vec{n}_\parallel$ and a component
perpendicular to $\vec{\theta}$ denoted by $\vec{n}_\perp$ (the index
$i$ is dropped for the sake of simplicity). These components are given
by
\begin{align*}
 \vec{n}_\parallel &= \hat\theta(\hat\theta\cdot\vec{n}),\ha 
 \hat \theta := \vec{\theta}/\|\vec{\theta}\|\\
 \vec{n}_\perp &= \vec{n} - \hat\theta(\hat\theta\cdot\vec{n}).
\end{align*}
Then the orientation update affects only $\vec{n}_\perp$ and the
updated $\vec{n}'$ is given by (with $\theta := \|\vec{\theta}\|$)
\begin{align}
  \label{rotation-update}
  \vec{n}_i' = \hat\theta(\hat\theta\cdot\vec{n}_i)
  (1 - \cos\theta) + \vec{n}_i\cos\theta + 
  \hat\theta\times \vec{n}_i \sin\theta, \ha i = 1,2,3.
\end{align}

\section{Reducing the systematic error in mean first passage time algorithm}
\label{appendix:error}

Applying the Euler algorithm \eq{langevin-euler} to a mean first
passage time problem gives rise to two sorts of errors.  First there
exists the statistical error, which is proportional to $1/\sqrt{N}$,
where $N$ is the number of iterations the algorithm is applied. The
extent of the statistical error of the measured mean value can be
calculated during the simulation. For the measurements performed in
sections \ref{sec:2d} and \ref{sec:3d} typically $N = 10^4-10^5$
iterations where chosen resulting in statistical errors in the range
of $<1\%$. Error-bars in these sections refer to the statistical
error.

The systematic error for the mean first passage time calculated by use
of an Euler algorithm scales with $\sqrt{\Delta t}$, although the
error of the particle position is only of the order of $\Delta t$
~\cite{honerkamp:94}. Thus to decrease the systematic error by a
factor of 10 one must increase the numerical cost by a factor of 100.
One way to obtain accurate results at moderate numerical cost is to
measure the mean first passage time for various intermediate numerical
time steps. Fitting these results to $a + b\sqrt{\Delta t}$ allows the
extrapolation to $\Delta t \rightarrow 0$.  \fig{error} shows an
example where this procedure was applied to the case of homogeneous
coverage as considered in \sec{sec:MFPT}. The resulting mean first
passage time then deviates by $0.2\%$ from the value obtained from
quadrature of \eq{z-mfpt}. This is the same accuracy as we have for
the implemented mobility functions themselves (cf. appendix
\ref{appendix:mobility}).

%\bibliography{misc,paper,textbooks,project1}
%\bibliographystyle{apsrev}

\clearpage

\begin{figure}
\caption{\label{fig:cartoon} Cartoon of a spherical particle with
      radius $R$ moving in linear shear flow above a wall. The height
      $z$ of the sphere center above the substrate obeys $z > R$. Bond
      formation between particle and wall is identified with spatial
      proximity between the receptor patches on the particle and the
      ligand patches on the wall being smaller than some prescribed
      encounter radius, that is overlap of the gray areas.}
\end{figure}

\begin{figure}
\caption{\label{falling} Falling sphere in shear flow. For different
      values of the shear rate (represented by the P\'eclet number
      $Pe$) and the driving force (represented by $f$ or $Pe_z = f\
      Pe$) the $z$-coordinate and the orientation angle $\theta$ are
      plotted versus the $x$-coordinate.}
\end{figure}

\begin{figure}
\caption{\label{fig-z-dist} Probability distribution function
      $\Psi(z,t)$ numerically obtained from $N = 10^5$ sample paths
      for ten consecutive points in time.  The initial distribution
      was $\Psi(z, t_0) = \delta(z - 3)$ at $t = t_0$, $Pe_z = 2$.}
\end{figure}

\begin{figure}
\caption{\label{mfpt} Results of first passage time simulations with
      encounter radius $r_0 = 10^{-3}$.  (a) Mean first passage time
      $T$ as a function of $Pe_z$ for different starting heights. Dots
      are the results from simulations with $N = 10^4$ runs and time
      step $\Delta t = 10^{-5}$.  Lines are the results from the
      quadrature of (\ref{z-mfpt}). (b) Distribution of first passage
      times for different values of $Pe_z$ (numerical parameters $N =
      10^5, \Delta t = 10^{-5}$.).}
\end{figure}

\begin{figure}
\caption{\label{initial_height} Mean first passage time dependence on
      the initial height $z_0$ in two dimensions. The sphere is
      covered with $N_r = 10$ receptor patches and the ligand density
      is $\rho_{l} = 0.01$.  We plot $\mean{T(z_0,\theta|C)}_{\theta,
      x}$ (+) and $\mean{T(z_0,\theta|C)}_{\theta, x} + T(z = 10|
      z_0)$ (x) as a function of $z_0$, where $ T(z = 10| z_0)$ is
      obtained from \eq{z-mfpt}. For $z_0 > 1 + r_0$ the latter curve
      is constant at the value $\mean{T(z=10,\theta|C)}_{\theta, x}$
      as predicted by the addition theorem \eq{mfpt:add}. (Numerical
      parameters: $N = 10^5, \Delta t = 10^{-5}$.)}
\end{figure}

\begin{figure}
\caption{\label{fig:rpatch} (a) Example of a sphere restricted to move
      in two dimensions and covered with $N_r=4$ receptor patches,
      which are regularly distributed over the circumference.  (b)
      Illustration of the range of $\theta$ in which encounter occurs.
      This range is given by $2\theta_0$ with $\theta_0(z) =
      \arccos(z/(1 + r_0))+ r_p$.  (c) The absorbing boundary $C$ in
      the $(z,\theta)$-plane is periodic with respect to $\theta$ with
      period $\theta_s = 2\pi/N_r$. For large numbers of receptor
      patches $\theta_s$ the different patches start to overlap. Then
      encounter is possible for all values of $\theta$.}
\end{figure}

\begin{figure}
\caption{\label{fig:receptorresults:one} The mean first passage time
      averaged over the initial orientation (log-log plots).  (a)
      Plotted as a function of $Pe$; different symbols refer to
      different numbers of receptor patches. (b) The mean first
      passage time is plotted as a function of the receptor density
      $\rho_r \propto N_r$ for different values of $Pe$.  (c)
      $\mean{T}_\theta$ as a function of $N_r$ in the diffusive regime
      ($Pe \approx 0$) for different values of the capture range
      $r_0$, but fixed value of cluster-size $r_p = 0.001$.  (d) The
      distribution of $\theta$-averaged first passage time is shown
      for $N_r = 5,20,50$ receptor patches.  (Numerical parameters for
      each data point: $N = 10^5, \Delta t = 10^{-5}$.)}
\end{figure}

\begin{figure}
\caption{\label{fig:ldsetup} (a) Illustration of the situation with a
      density of receptor patches $\rho_r$ as well as a density of
      ligands $\rho_l$. The first passage time is now determined by an
      overlap of a receptor patch with a ligand patch. (b)
      $\mean{T}_{\theta, x}$ as function of the P\'eclet number $Pe$
      and the ligand density $\rho_l$ for different values of $N_r$
      (numerical parameters: $\Delta t = 5\cdot 10^{-6}, N = 10^4$).}
\end{figure}

\begin{figure}
\caption{\label{fig:ligandresults} (a) $\mean{T}_{\theta, x}$ is shown
    in the diffusion limit at $Pe \approx 0$ as a function the ligand
    density $\rho_l$.  Inset (plot for $\rho_r \approx 1$): The mean
    first passage time scales as $\mean{T}_{\theta, x} \propto
    1/\rho_l^2$ (numerical parameters: $\Delta t = 10^{-5}, N =
    10^{5}$).  (b) Dependence of $\mean{T}_{\theta, x}$ in the
    diffusive limit at $Pe \approx 0$ on $\rho_r, \rho_l$, where
    $\rho_r$ has been varied by changing $N_r$ at fixed $r_p$
    (numerical parameters: $\Delta t = 10^{-5}, N = 10^{5}$).}
\end{figure}

\begin{figure}
\caption{\label{fig:3d} (a) The mean time for a receptor to first
      reach a wall homogeneously covered with ligands
      $\mean{T}_{\vec{\theta}}$ was calculated as a function of the
      P\'eclet number $Pe$.  (b) The dependence of the MFPT on the
      number of receptor patches $N_r$ for different values of the
      capture radius $r_0$.  Lines show the scaling with $1/N_r$.  (c)
      Dependence of $\mean{T}_{\vec{\theta}, x, y}$ on the 2D ligand
      density $\rho_l$ in the diffusive limit $Pe \approx 0$. For
      $\rho_l \ll 1$ the mean first passage time is proportional to
      $1/\rho_l$ (dotted lines). In the inset are plotted the mutual
      ratios of the averaged mean first passage times for $N_r =
      20,30,70$, showing that the dependence on the ligand density is
      nearly independent on the number of receptor patches $N_r$
      (numerical parameters: $N = 10^5, \Delta t = 5\cdot 10^{-5}, r_p
      = 10^{-3}$, $r_0 = 10^{-3}$ for (a); $r_0 = r_d = 10^{-2}$ for
      (c)).}
\end{figure}

\begin{figure}
\caption{\label{fig:3d:02} (a, b) Dependence of
      $\mean{T}_{\vec{\theta}}$ on the receptor patch radius $r_p$
      ($Pe \approx 0$).  The dotted lines are fits of $a/(b + r_p)$ to
      the simulation results.  (a) $r_0 = 0.001$, (b) $r_0 = 0.01$
      (numerical parameters: $N = 1-3 \cdot 10^5, \Delta t = 5\cdot
      10^{-5}$).  (c) For $N_r = 30$ the dependence on $r_p$ is shown
      for different values of the capture radius $r_0$. For better
      comparison the $r_0$-dependent part of the MFPT as given by
      \eq{z-mfpt} was subtracted.}
\end{figure}

\begin{figure}
\caption{\label{abbmobility} Dimensionless scalar mobility
      functions. On the left the functions are plotted vs. the
      dimensionless parameter $t$. On the right the functions are
      plotted vs. $1 - t$, thus better illustrating the asymptotic
      behavior for $t \rightarrow 1$.}
\end{figure}

\begin{figure}
\caption{\label{error} The mean first passage times for $Pe_z = 100$,
  $z_1 = 1.001, z_0 = 2$ as a function of the numerical time step. The
  points are the results from simulation experiments (error-bars
  denote their statistical error) with $N = 10^5$ iterations. The full
  line is a fit to $a + b\sqrt{\Delta t}$ using the {\sf gnuplot}
  implementation of the nonlinear least-squares (NLLS)
  Marquardt-Levenberg algorithm.  Extrapolating the fit to $\Delta t
  \rightarrow 0$ reduced the systematic error due to the finite time
  step.}
\end{figure}

\clearpage

\resizebox{.96\linewidth}{!}{\includegraphics{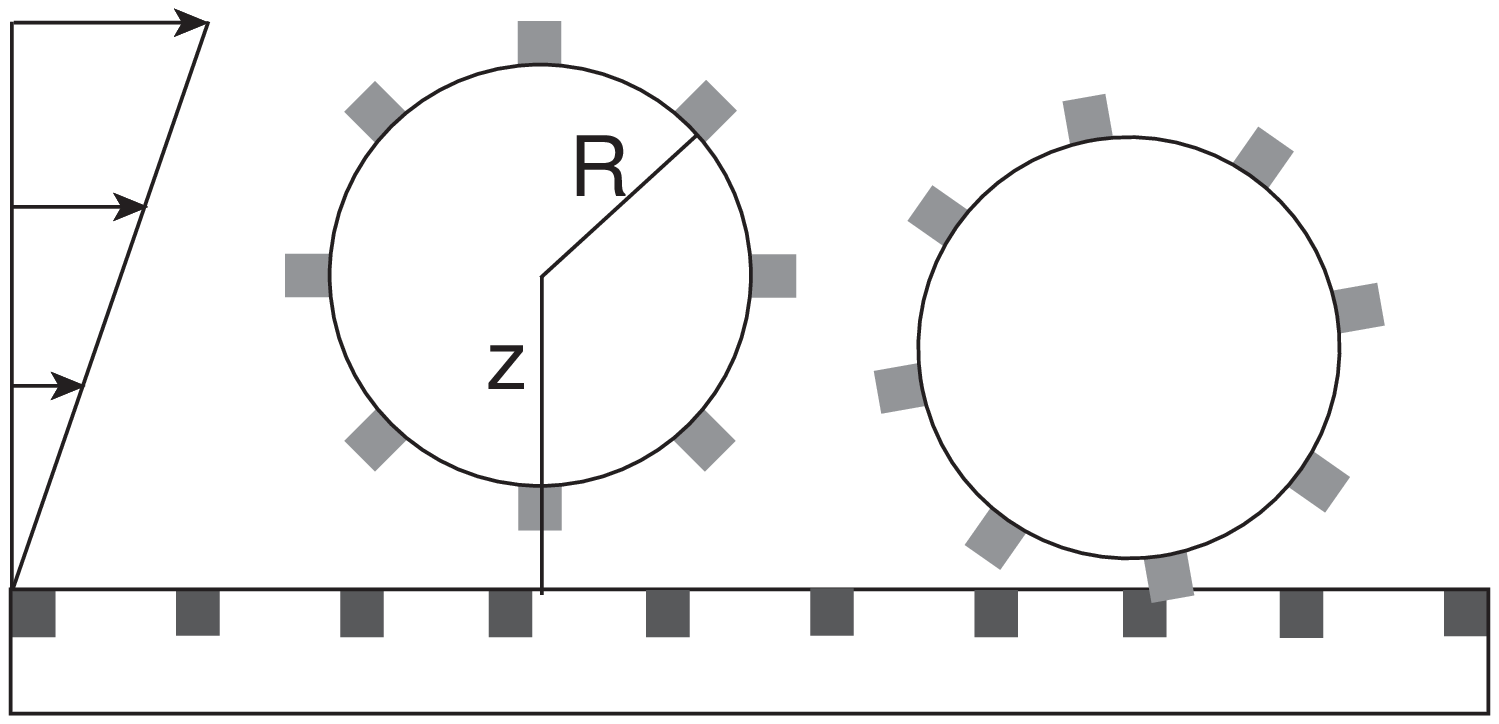}}

\fig{fig:cartoon}
\newpage

\resizebox{.96\linewidth}{!}{\includegraphics{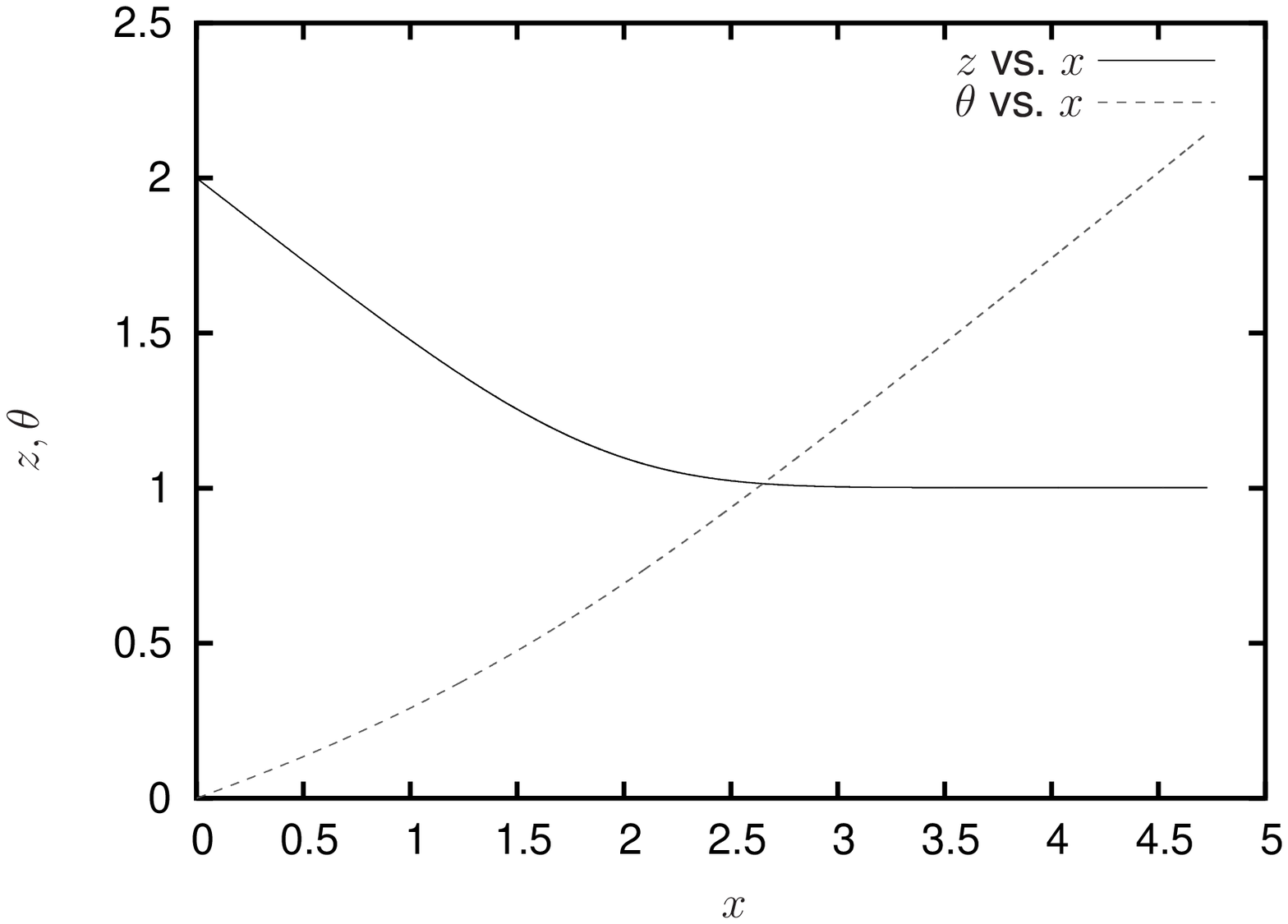}}

\fig{falling}a
\newpage

\resizebox{.96\linewidth}{!}{\includegraphics{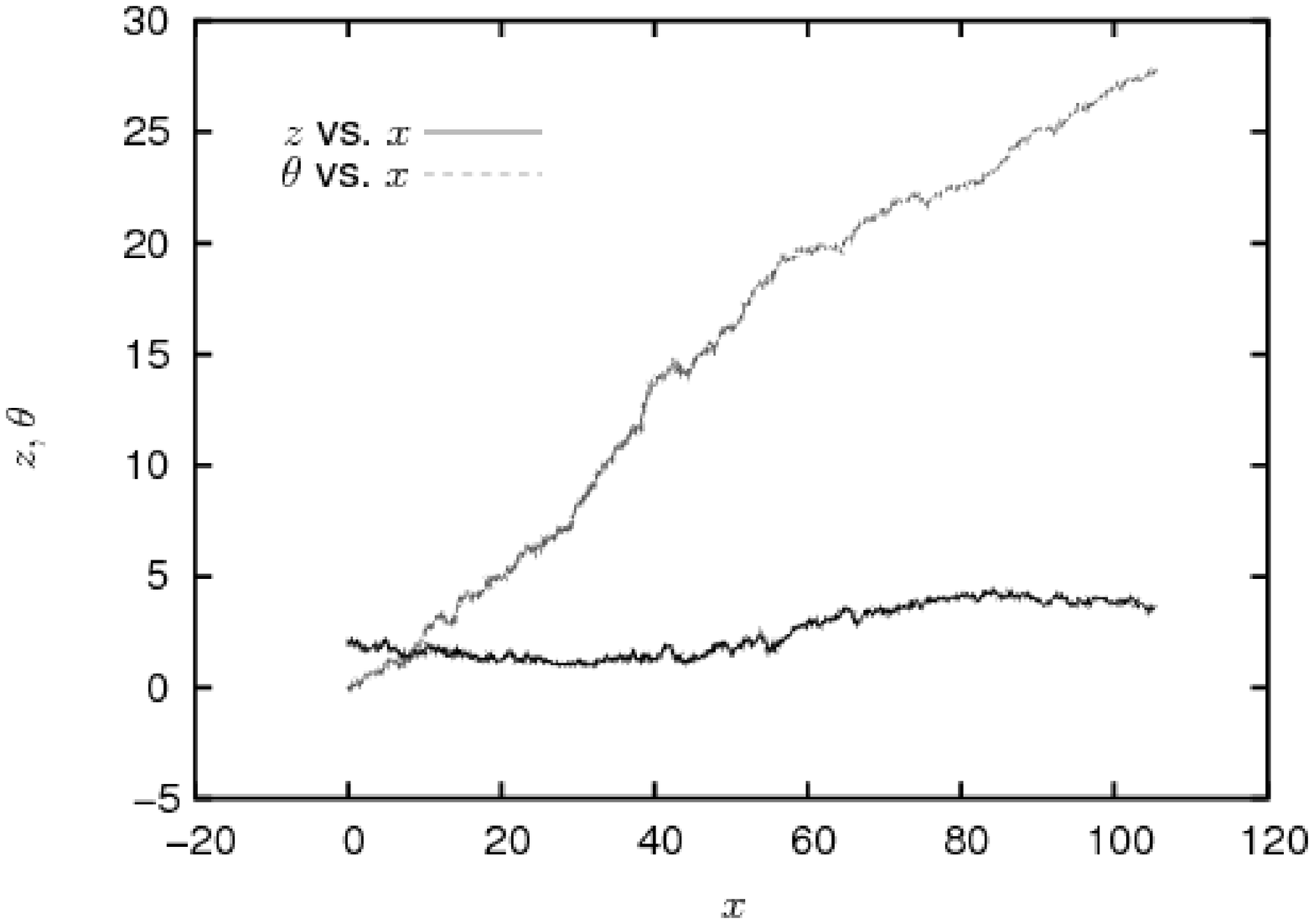}}

\fig{falling}b
\newpage

\resizebox{.96\linewidth}{!}{\includegraphics{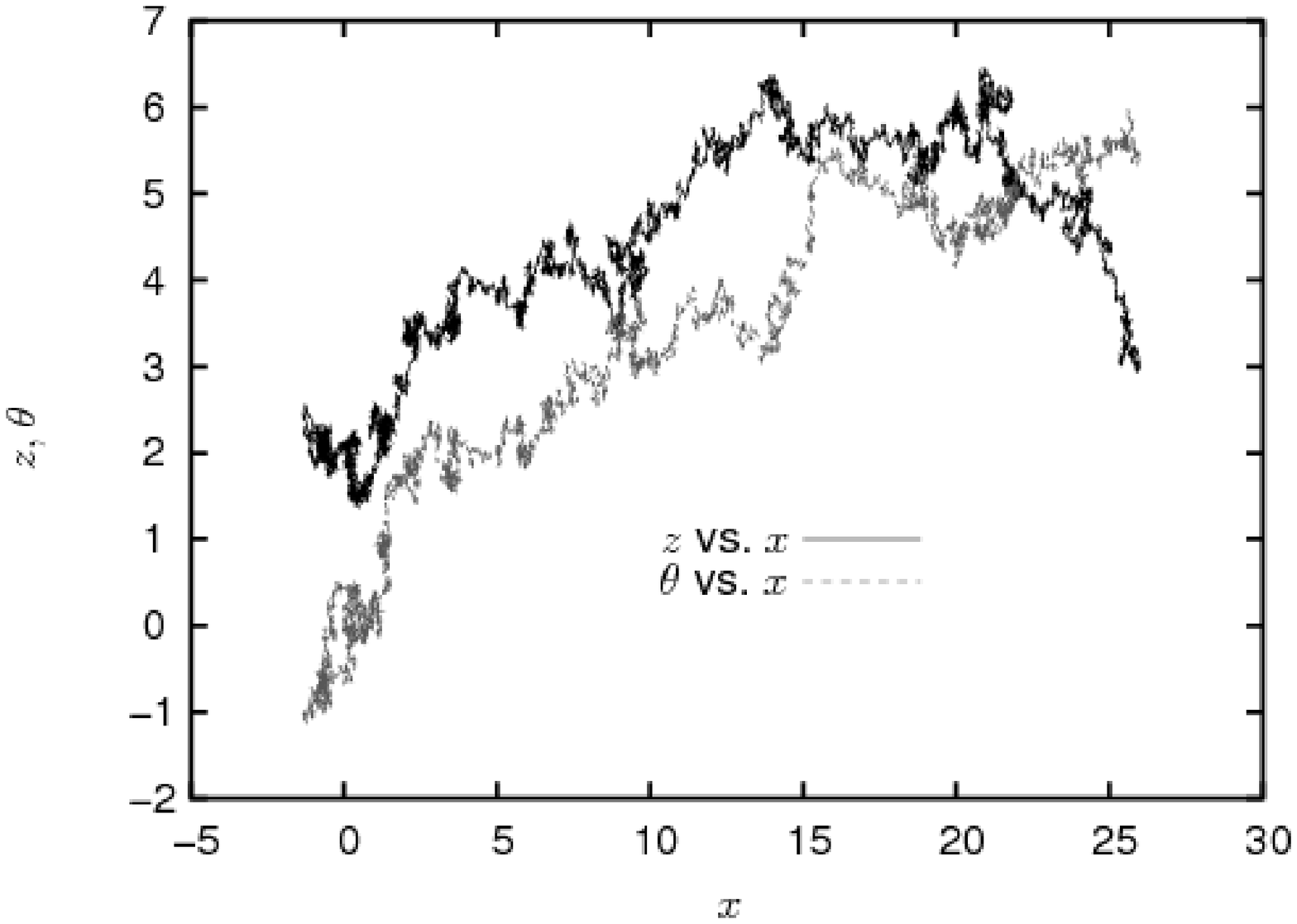}}

\fig{falling}c
\newpage

\resizebox{.96\linewidth}{!}{\includegraphics{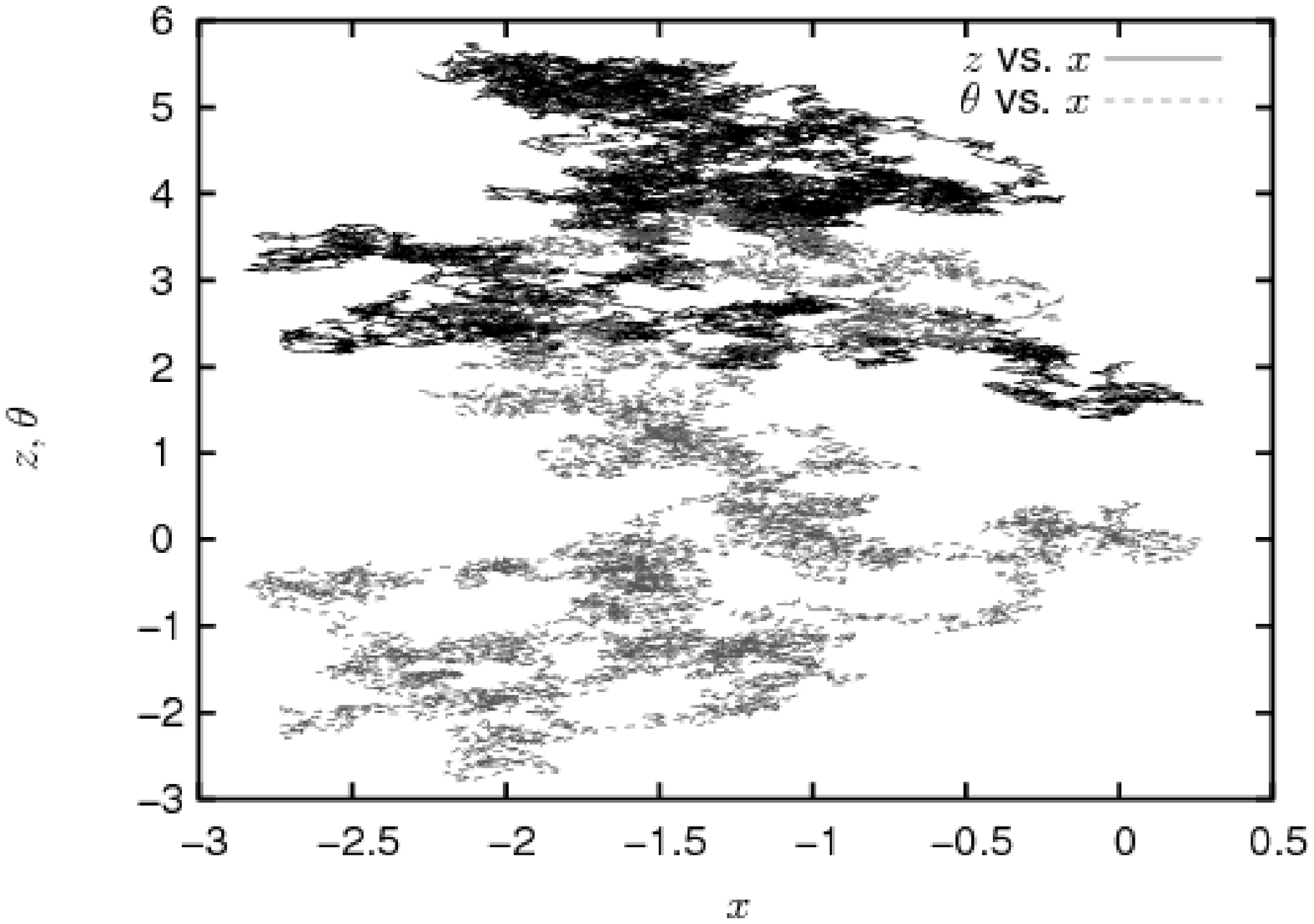}}

\fig{falling}d
\newpage

\resizebox{.96\linewidth}{!}{\includegraphics{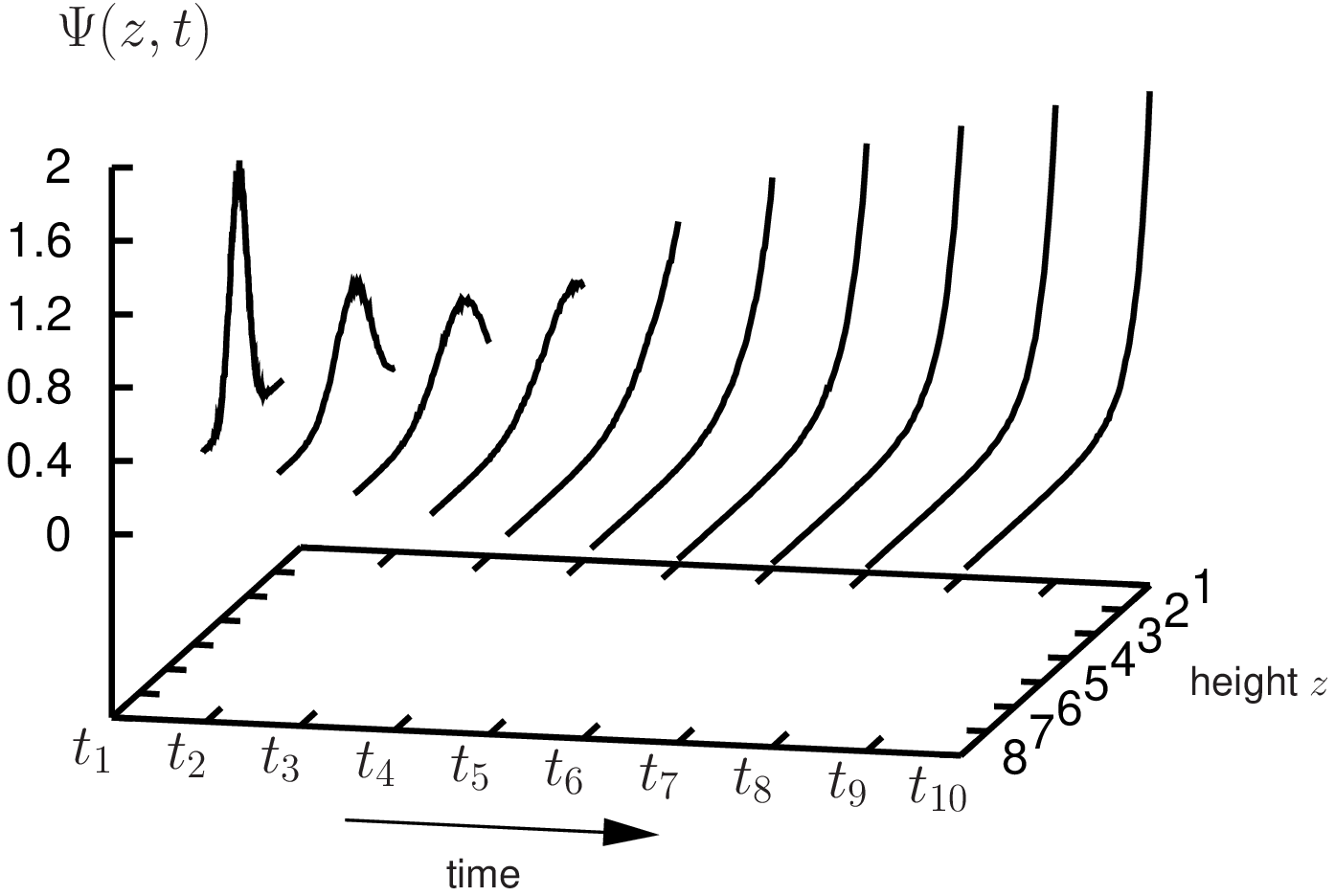}}

\fig{fig-z-dist}
\newpage
   
\resizebox{.96\linewidth}{!}{\includegraphics{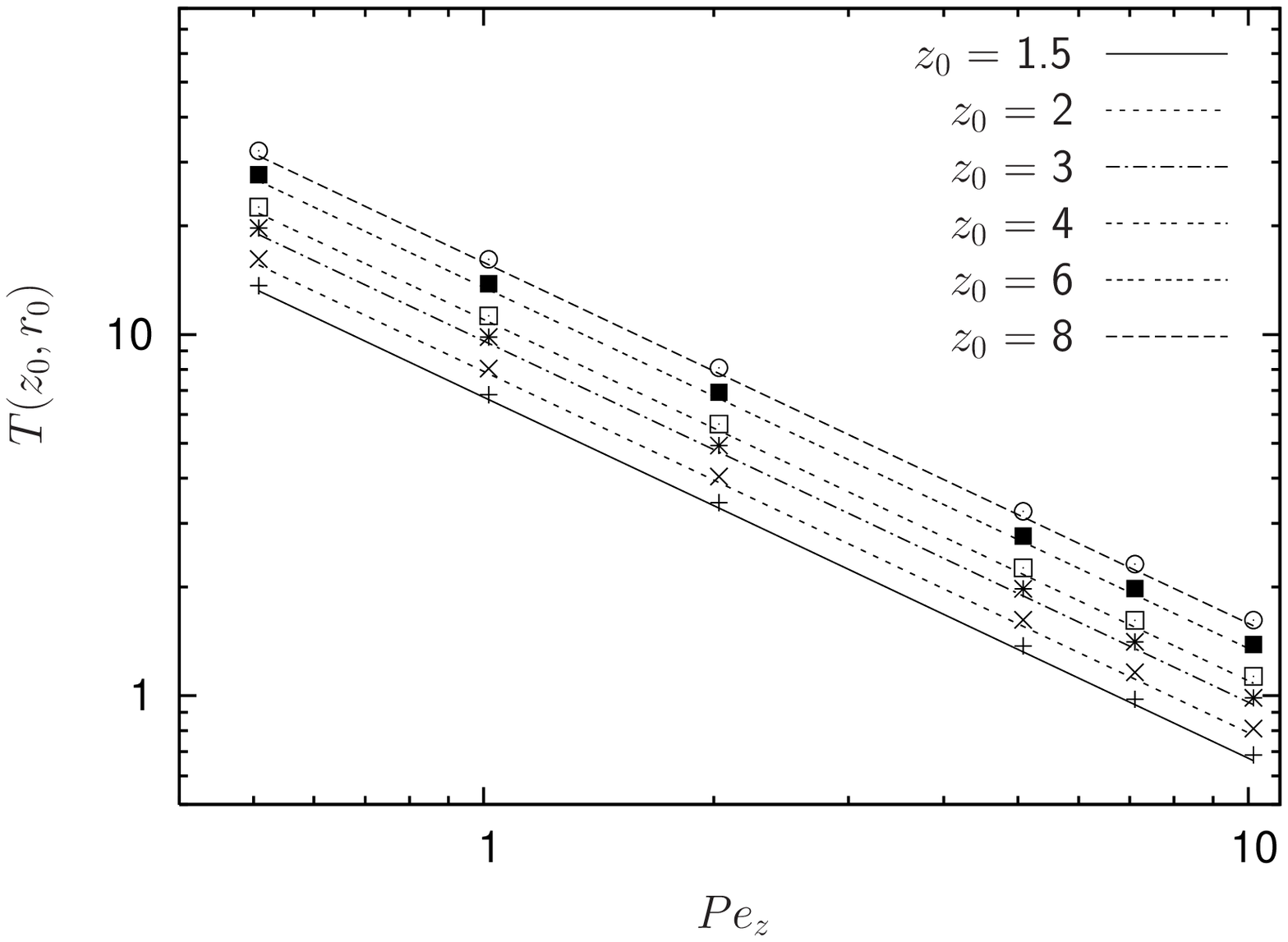}}

\fig{mfpt}a
\newpage

\resizebox{.96\linewidth}{!}{\includegraphics{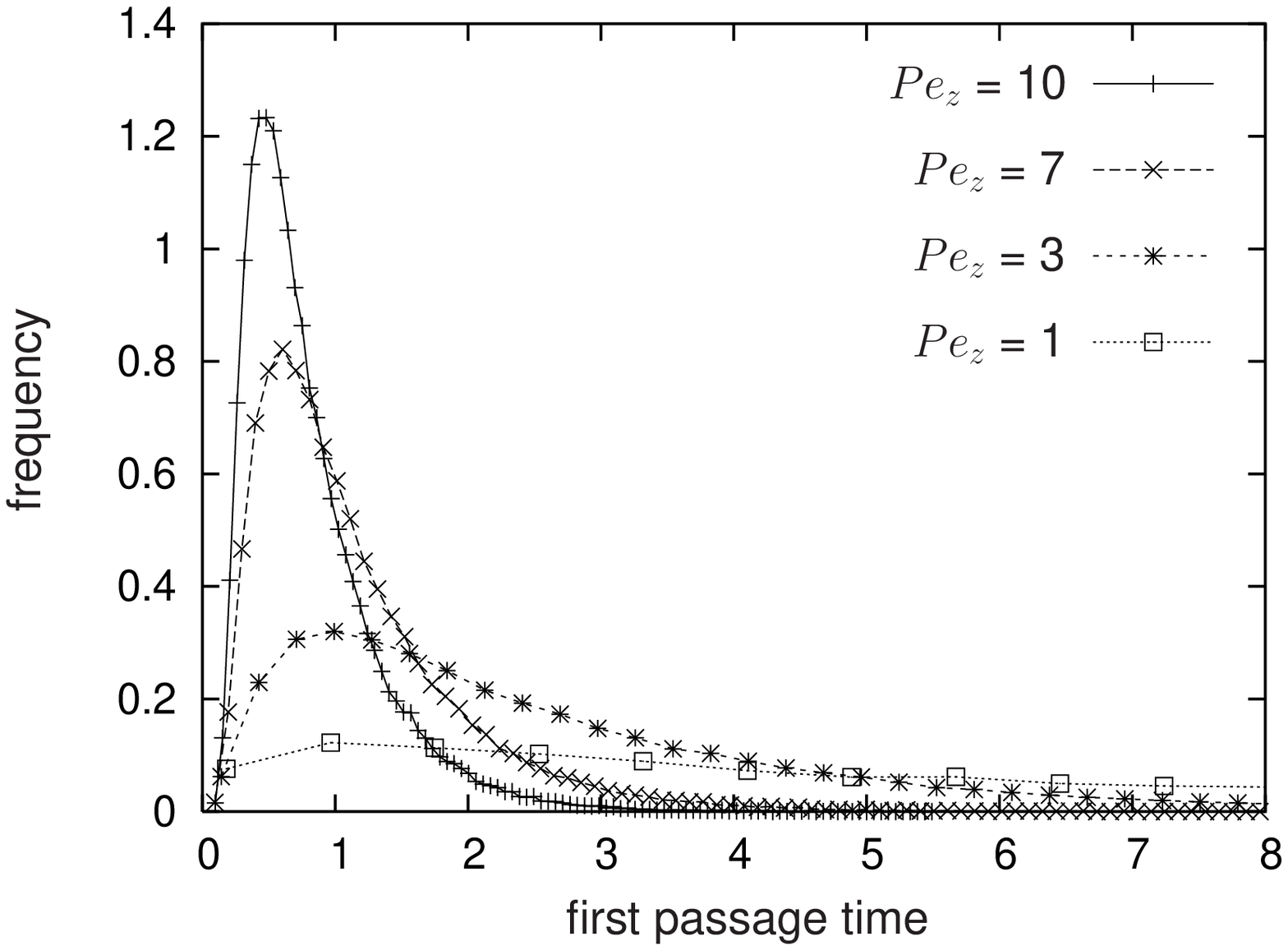}}

\fig{mfpt}b
\newpage

\resizebox{.96\linewidth}{!}{\includegraphics[scale=.424]{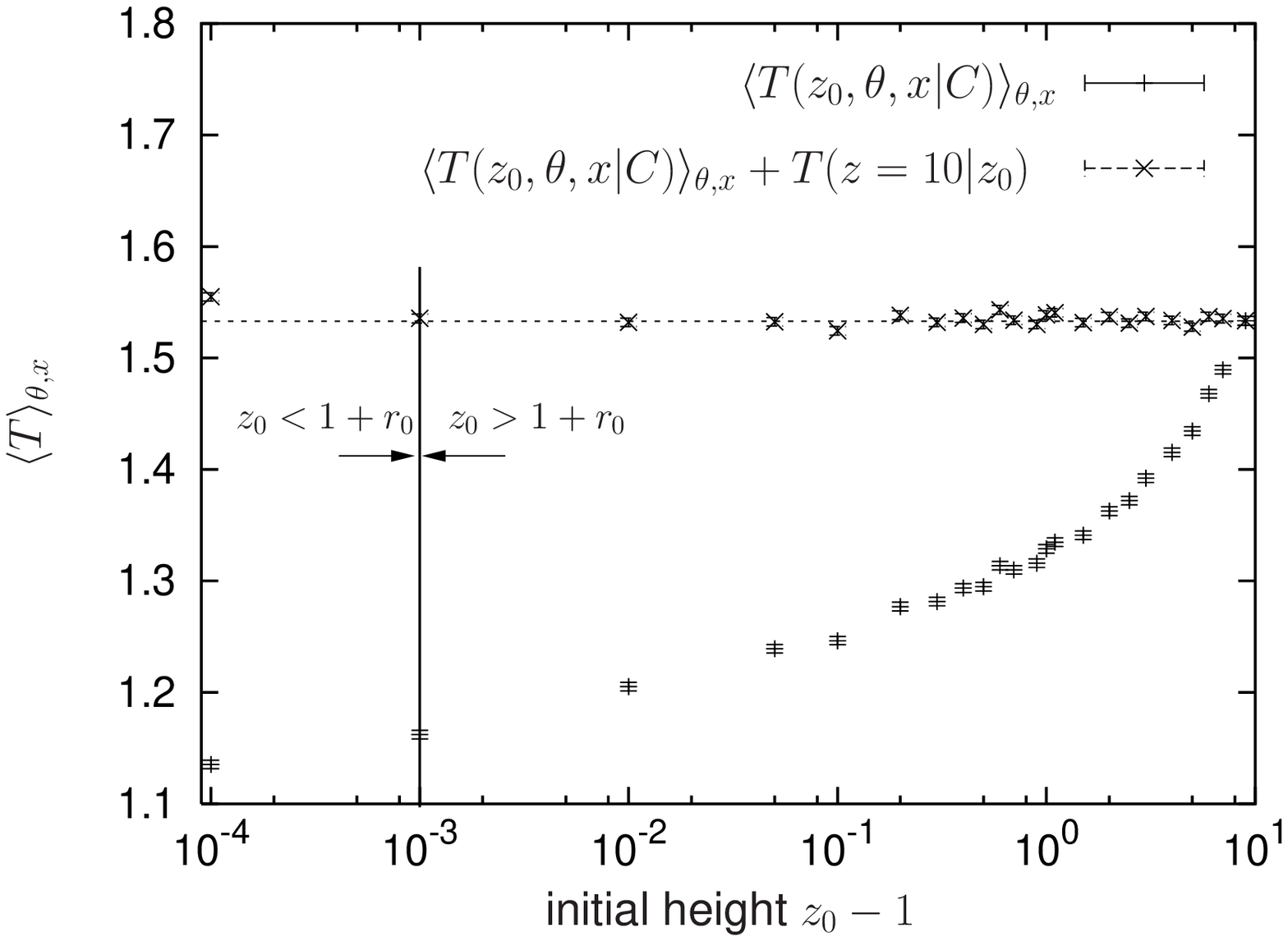}}

\fig{initial_height}
\newpage

\resizebox{.96\linewidth}{!}{\includegraphics{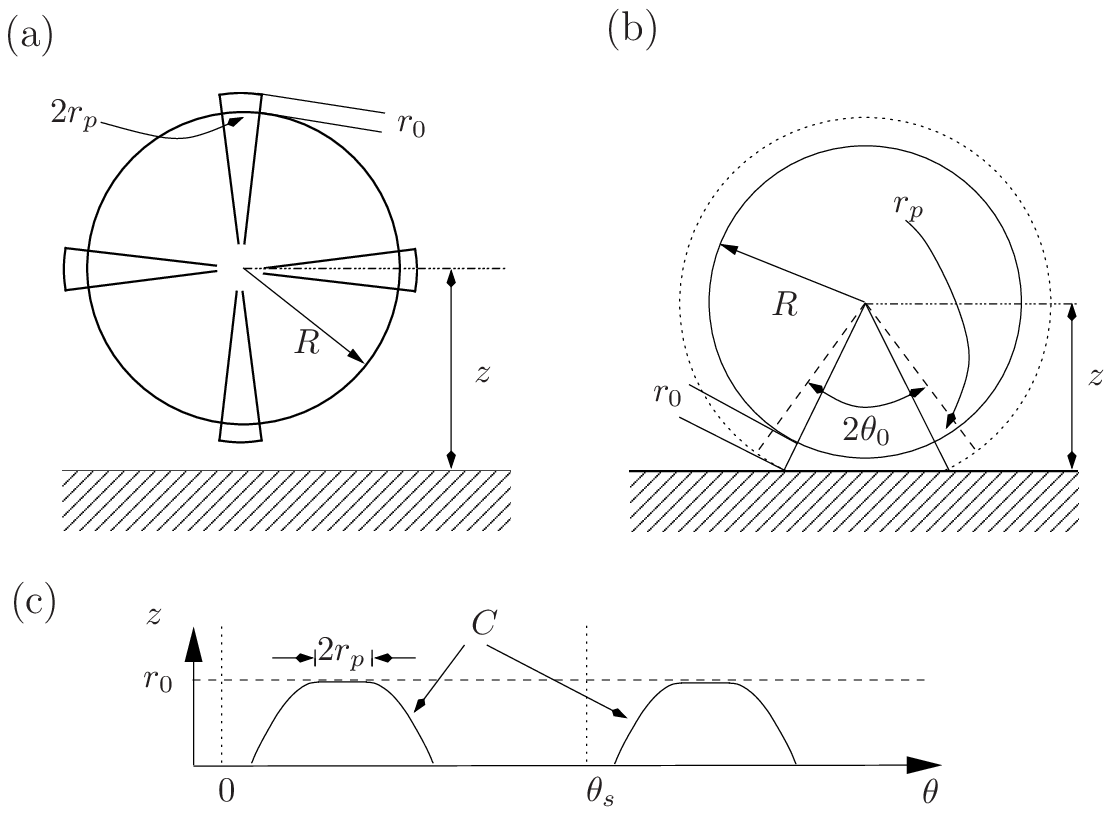}}

\fig{fig:rpatch}
\newpage

\resizebox{.96\linewidth}{!}{\includegraphics{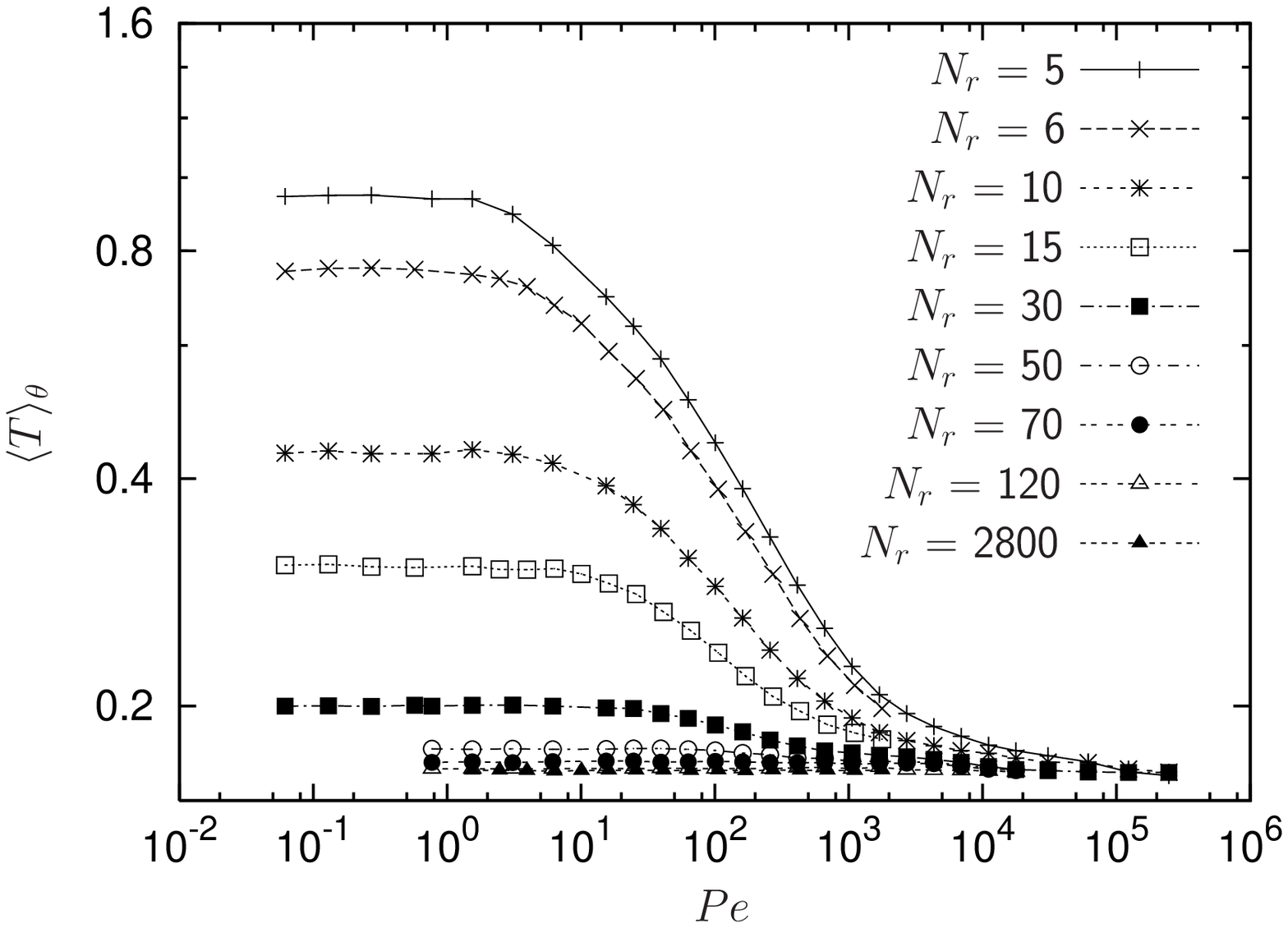}} 

\fig{fig:receptorresults:one}a
\newpage

\resizebox{.96\linewidth}{!}{\includegraphics{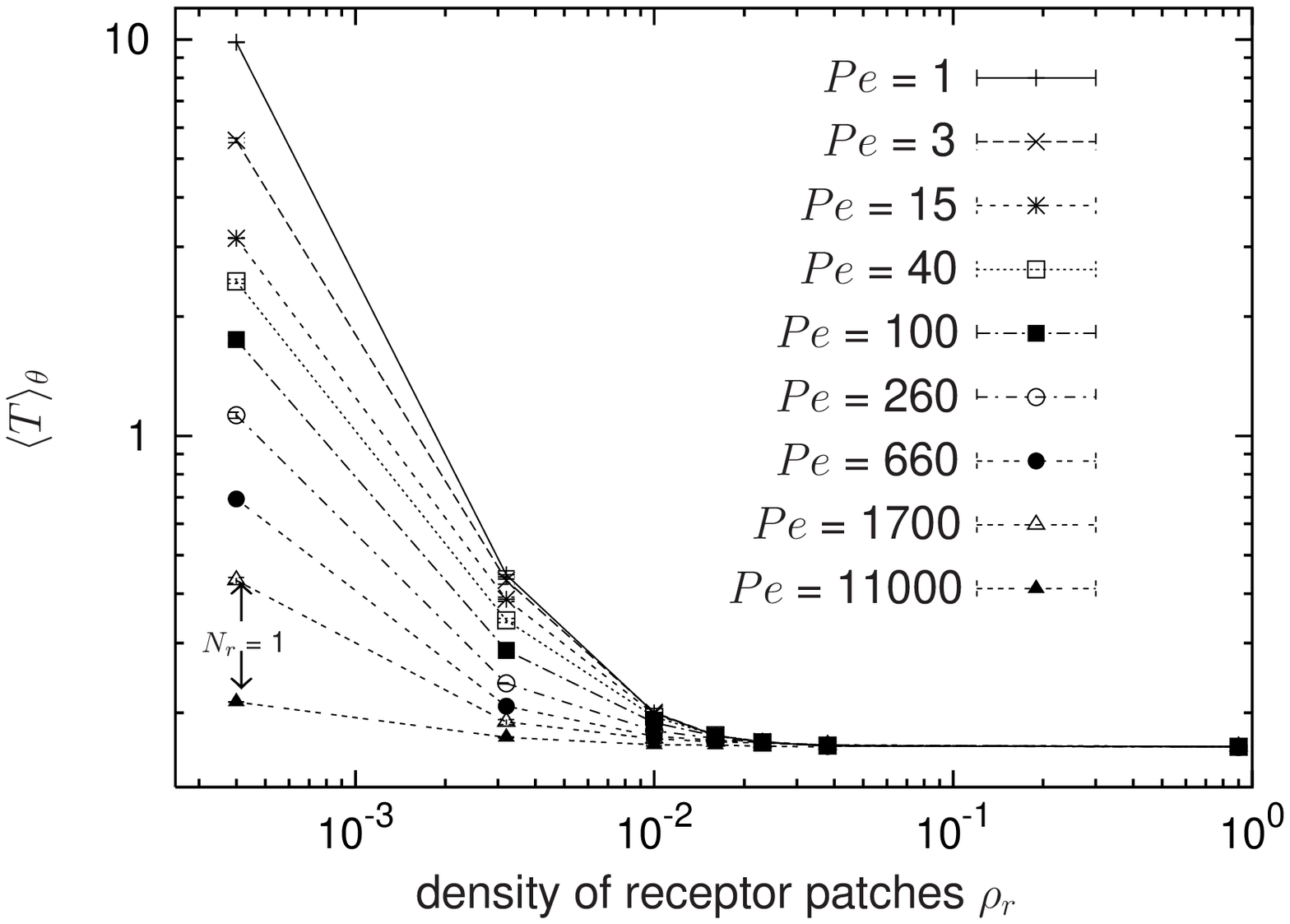}}

\fig{fig:receptorresults:one}b
\newpage

\resizebox{.96\linewidth}{!}{\includegraphics{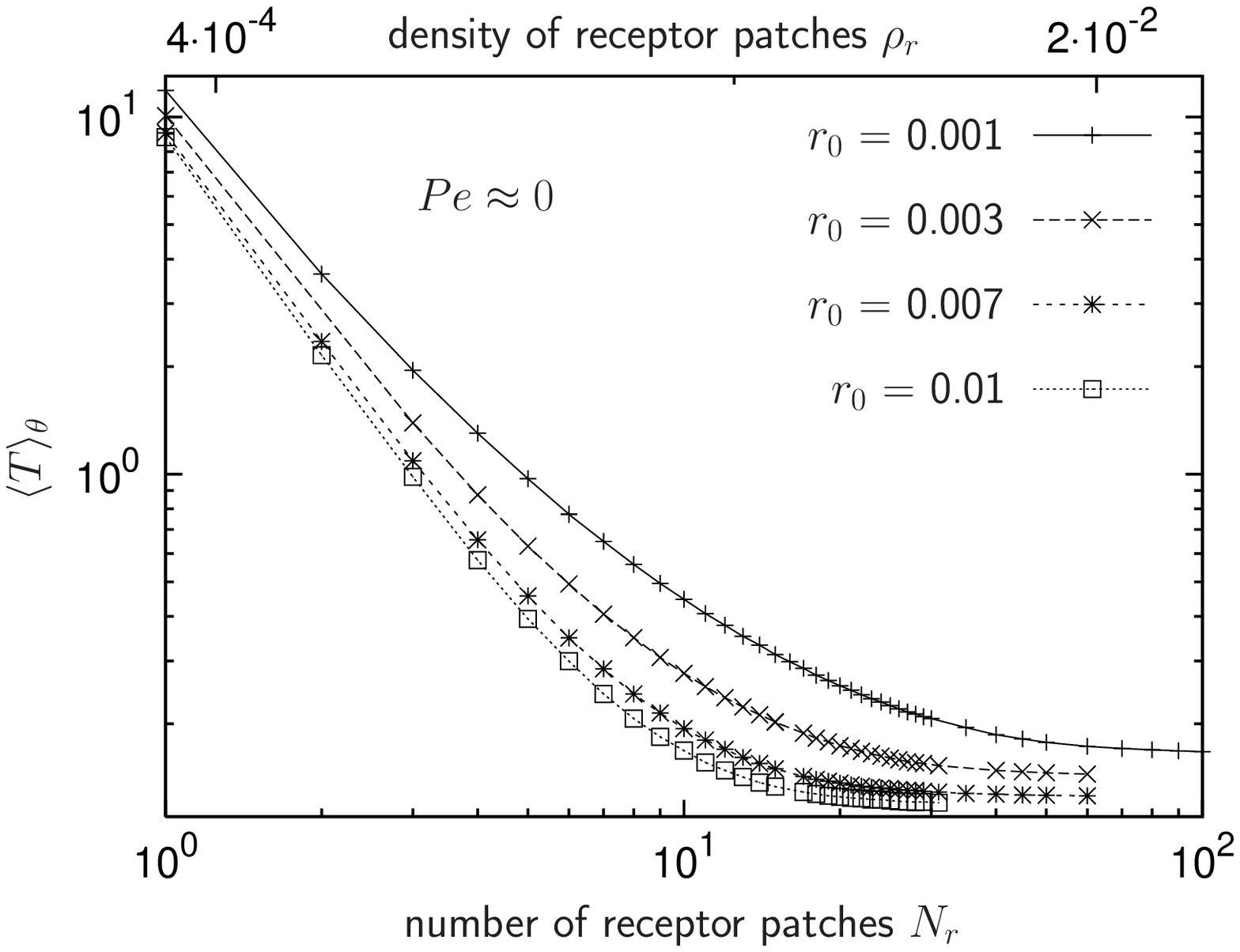}} 

\fig{fig:receptorresults:one}c
\newpage

\resizebox{.96\linewidth}{!}{\includegraphics{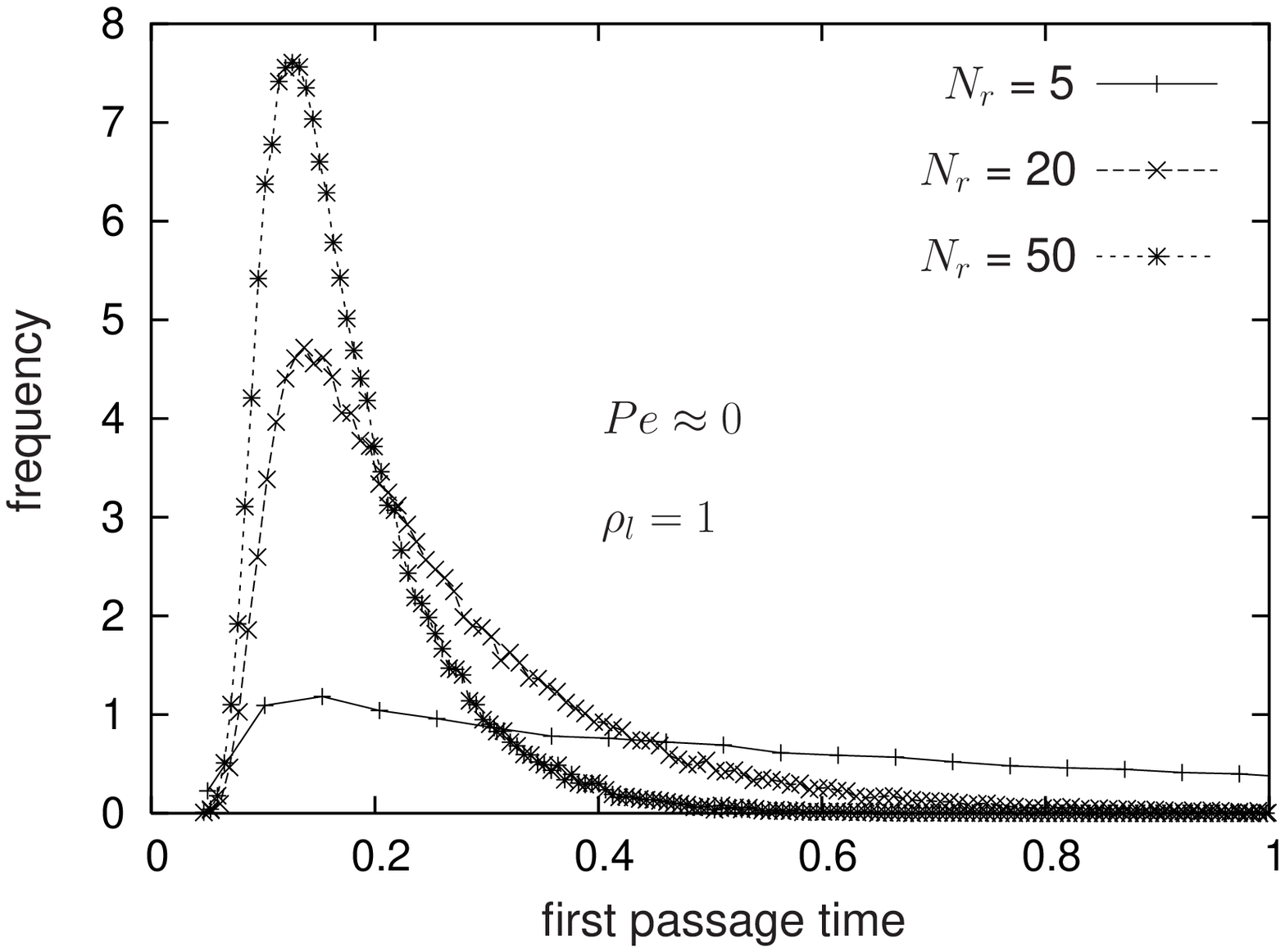}}

\fig{fig:receptorresults:one}d
\newpage

\resizebox{.96\linewidth}{!}{\includegraphics{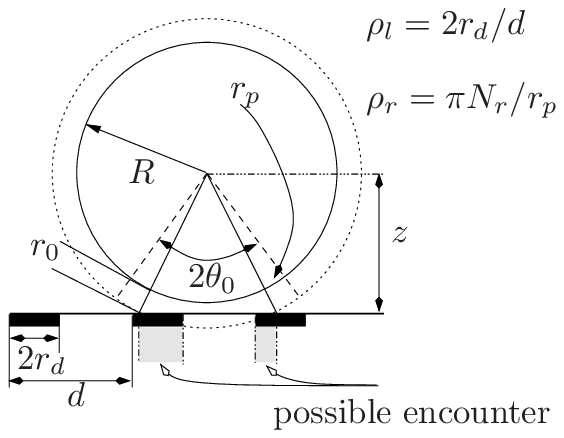}}   

\fig{fig:ldsetup}a
\newpage

\resizebox{.96\linewidth}{!}{\includegraphics{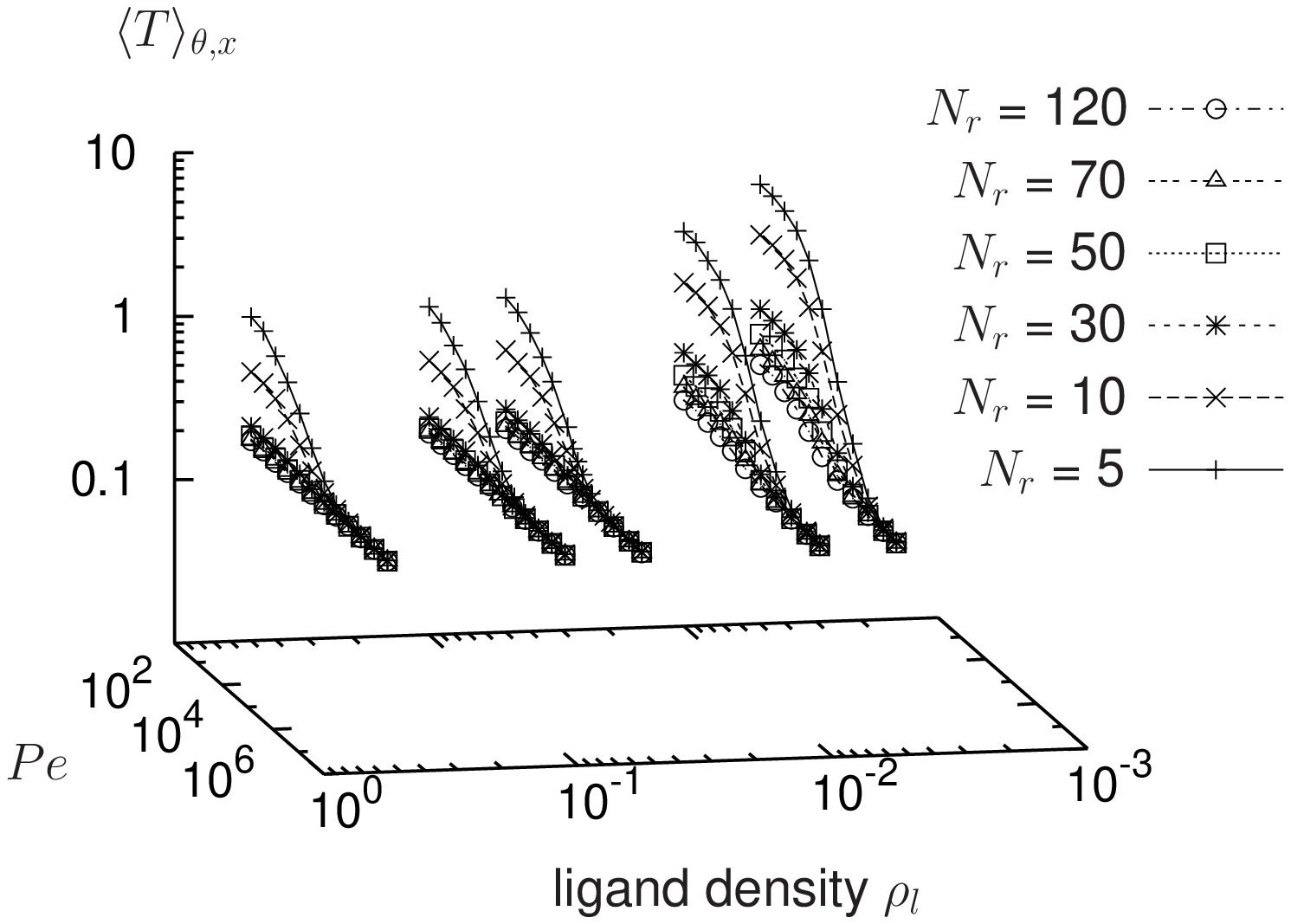}}

\fig{fig:ldsetup}b
\newpage

\resizebox{.96\linewidth}{!}{\includegraphics{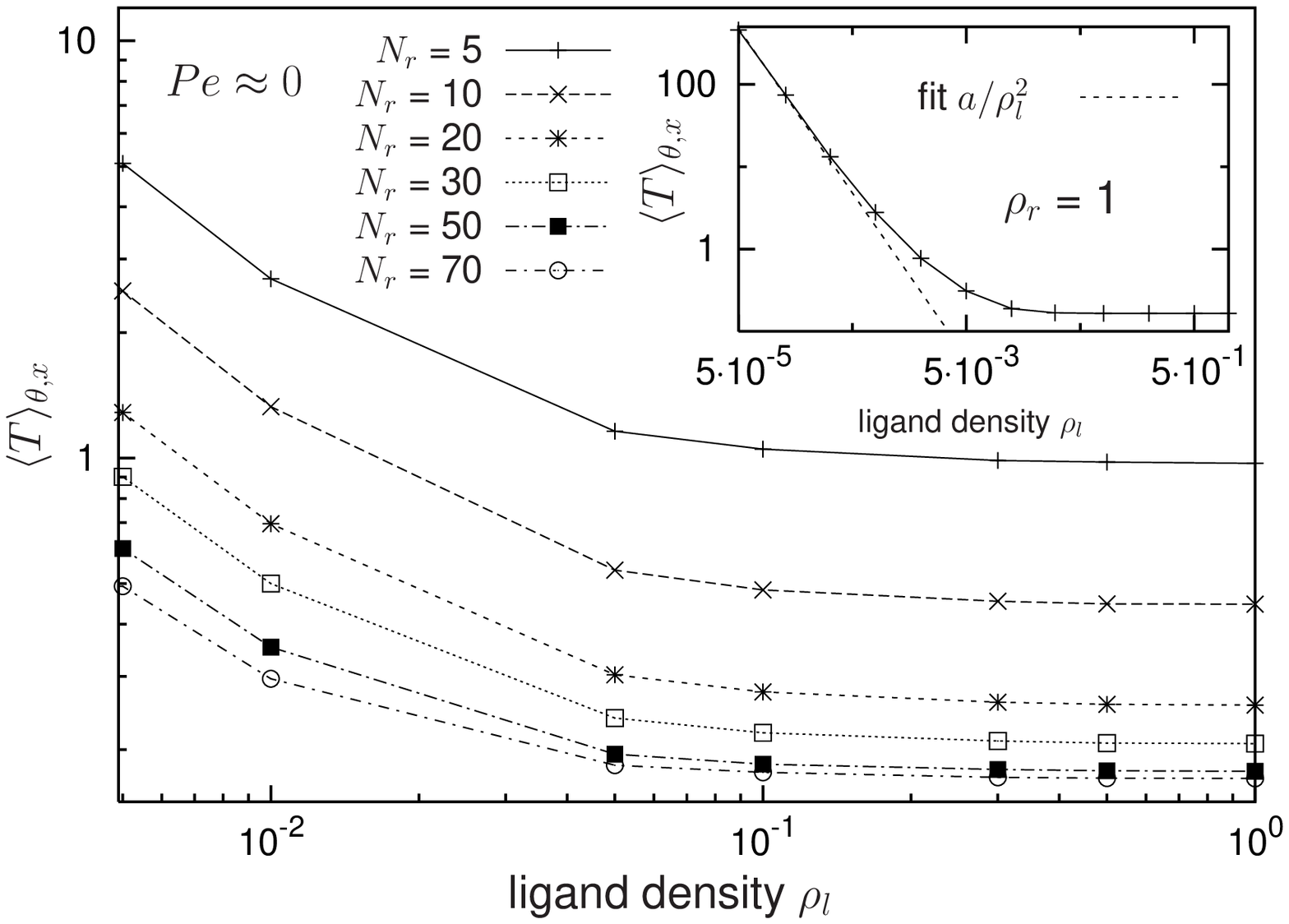}}

\fig{fig:ligandresults}
\newpage

\resizebox{.96\linewidth}{!}{\includegraphics{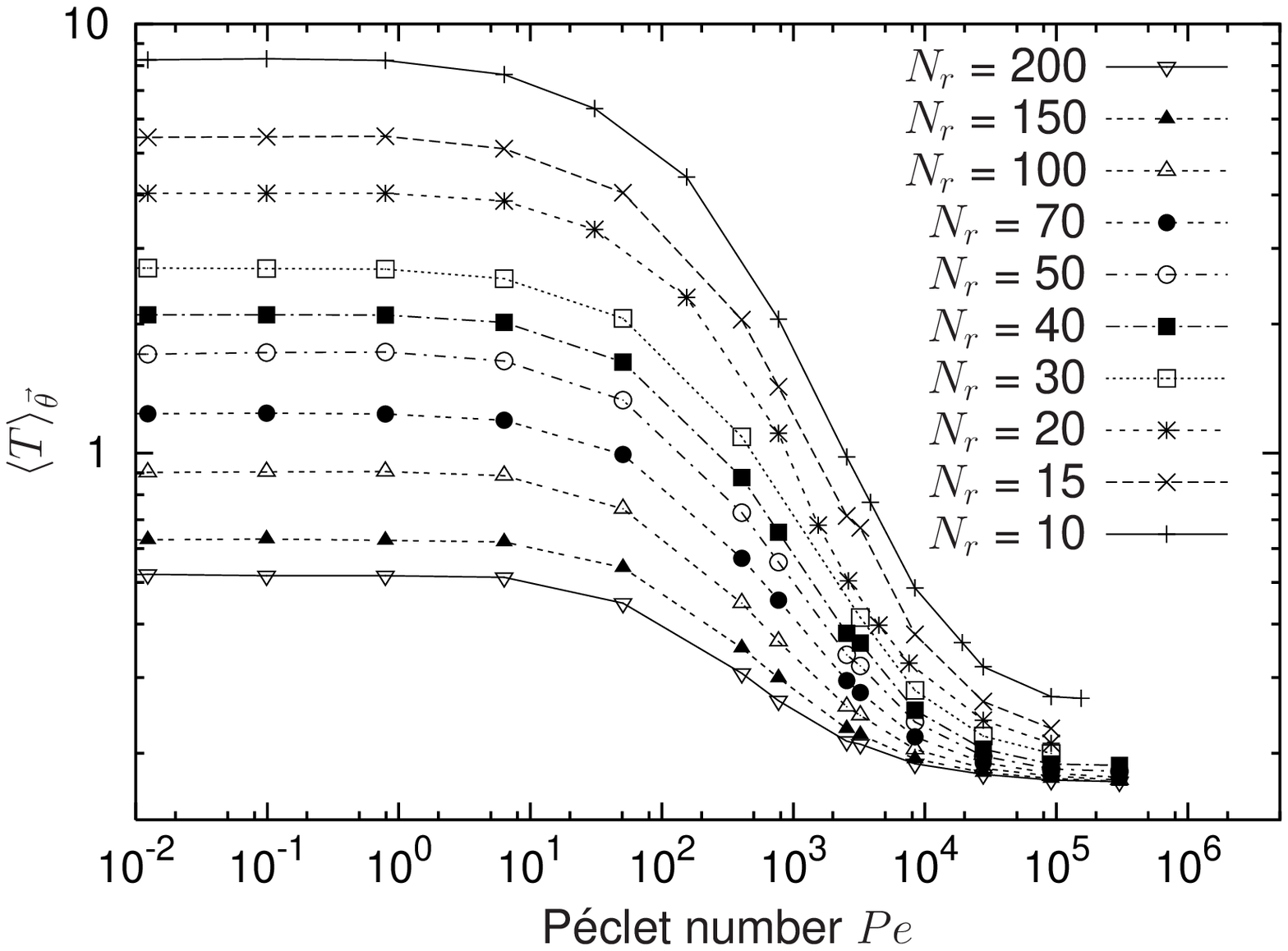}}

\fig{fig:3d}a
\newpage

\resizebox{.96\linewidth}{!}{\includegraphics{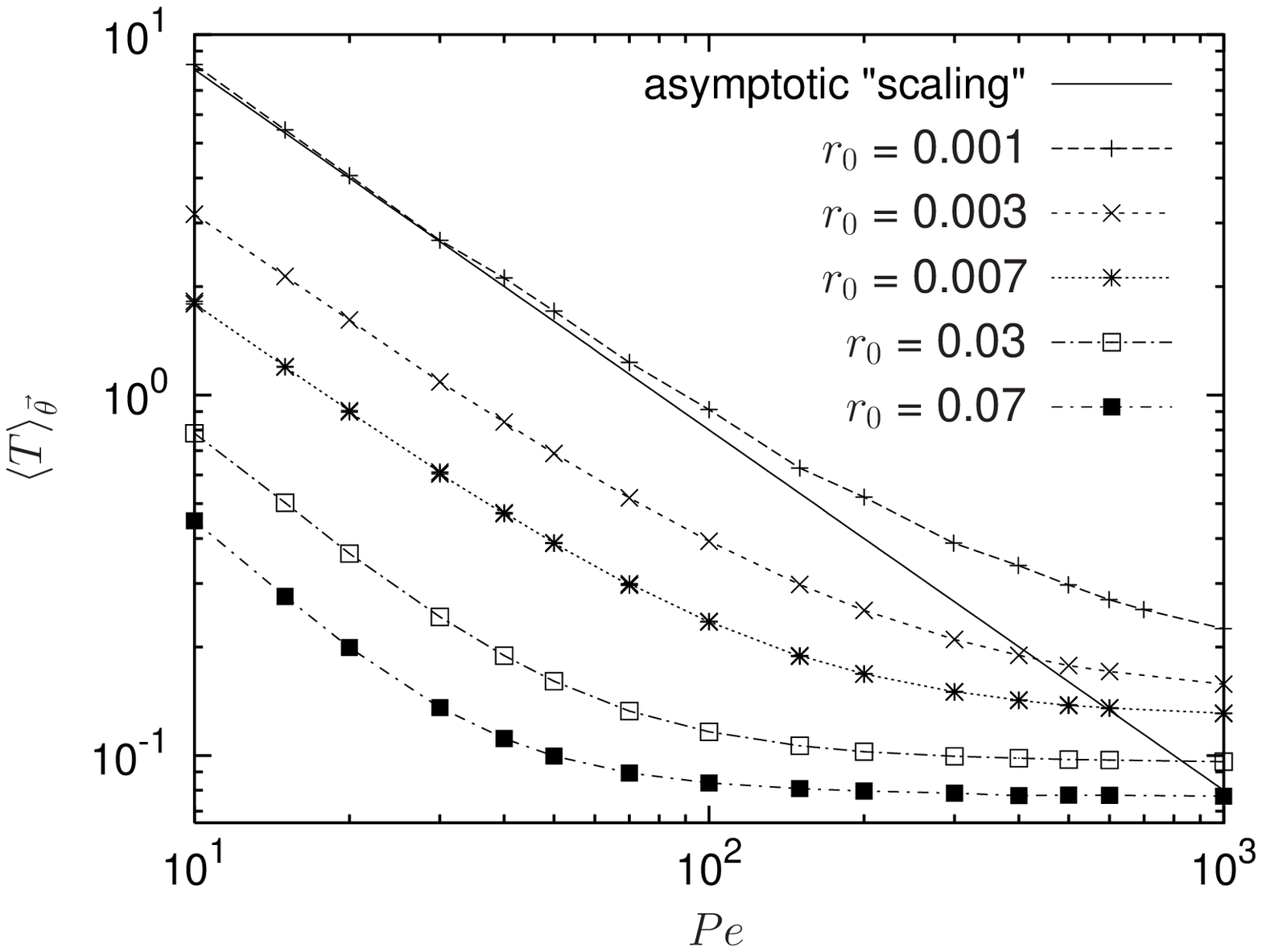}}

\fig{fig:3d}b
\newpage

\resizebox{.96\linewidth}{!}{\includegraphics{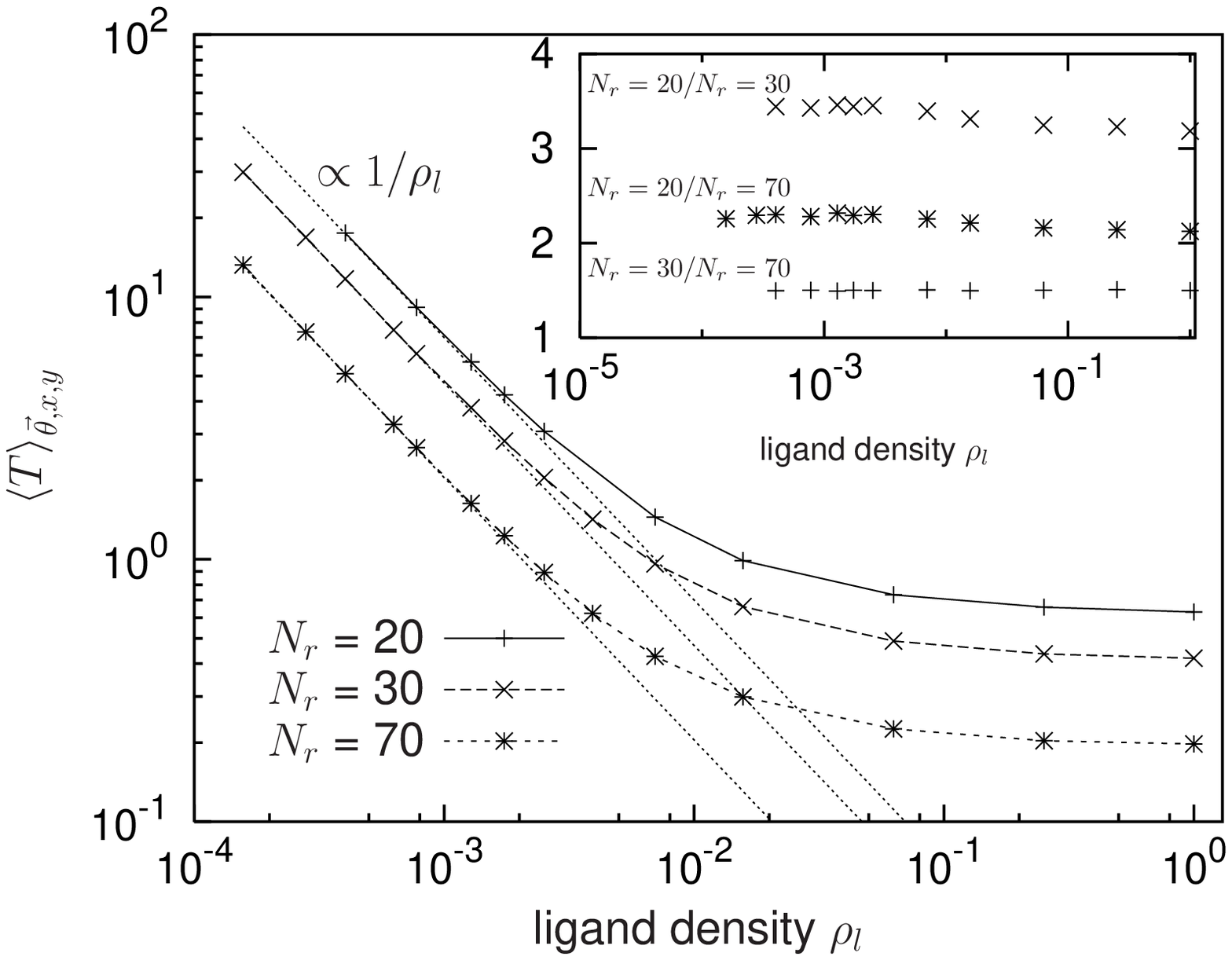}}

\fig{fig:3d}c
\newpage
	
\resizebox{.96\linewidth}{!}{\includegraphics{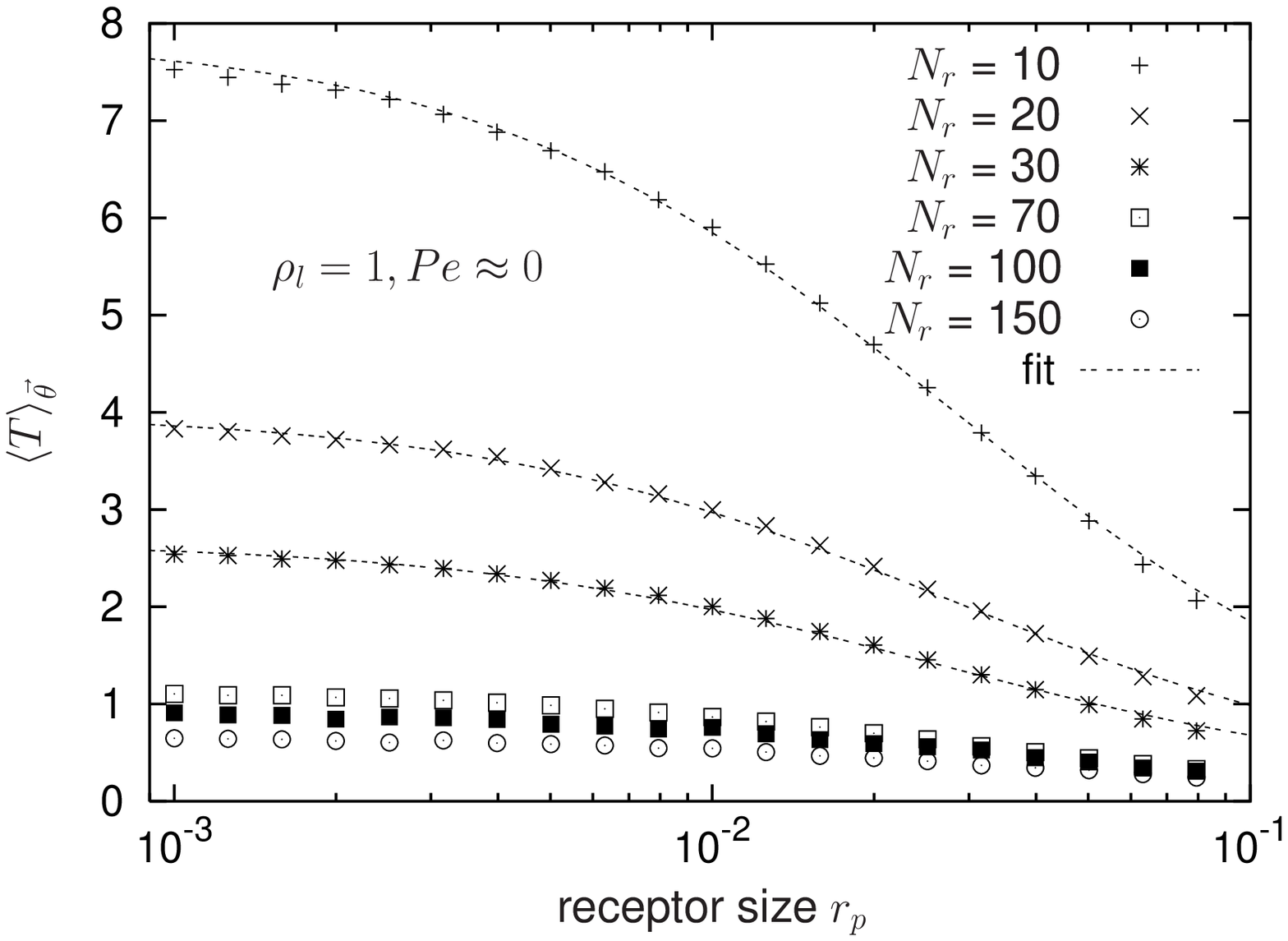}}

\fig{fig:3d:02}a
\newpage

\resizebox{.96\linewidth}{!}{\includegraphics{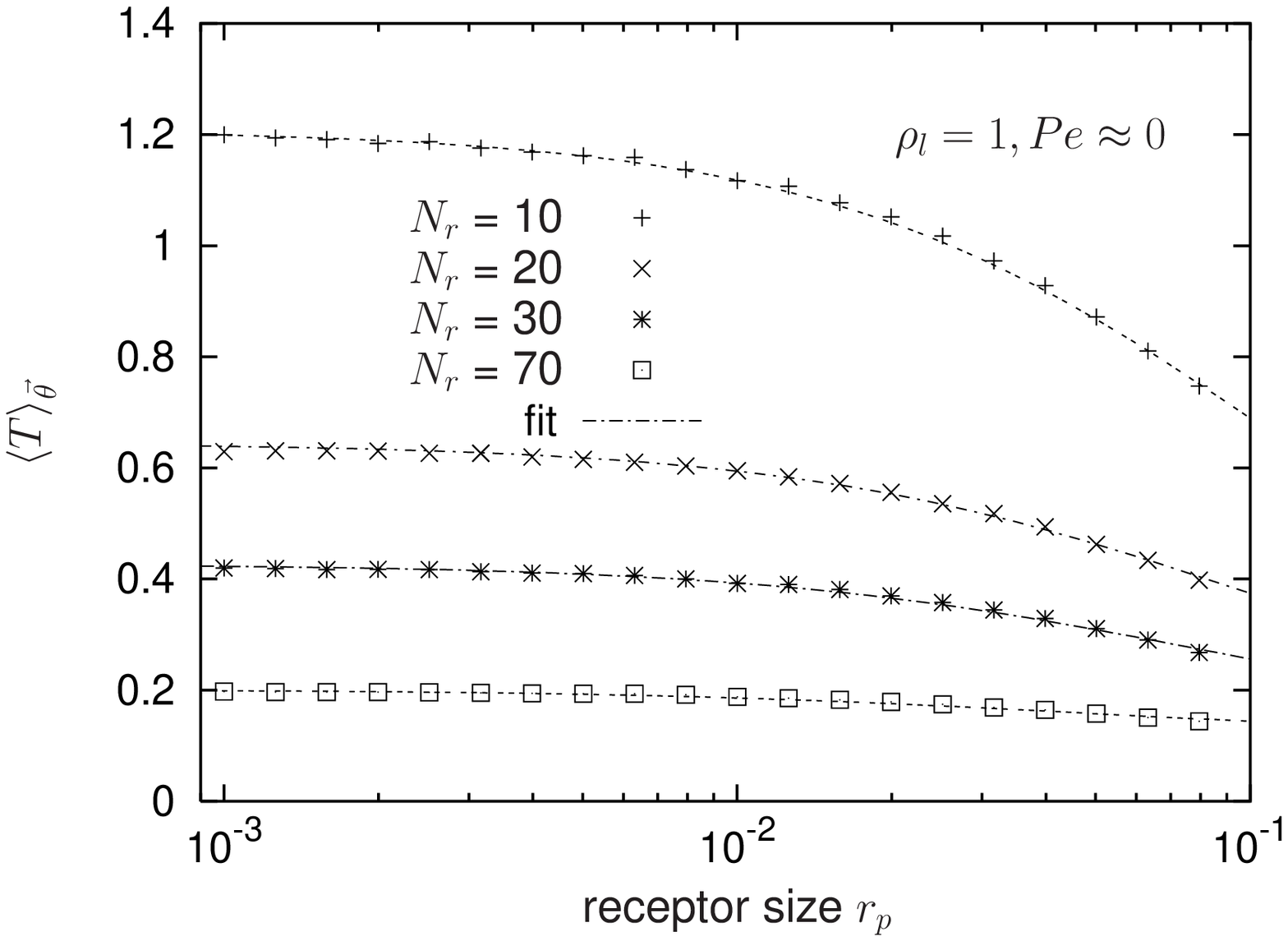}}

\fig{fig:3d:02}b
\newpage

\resizebox{.96\linewidth}{!}{\includegraphics{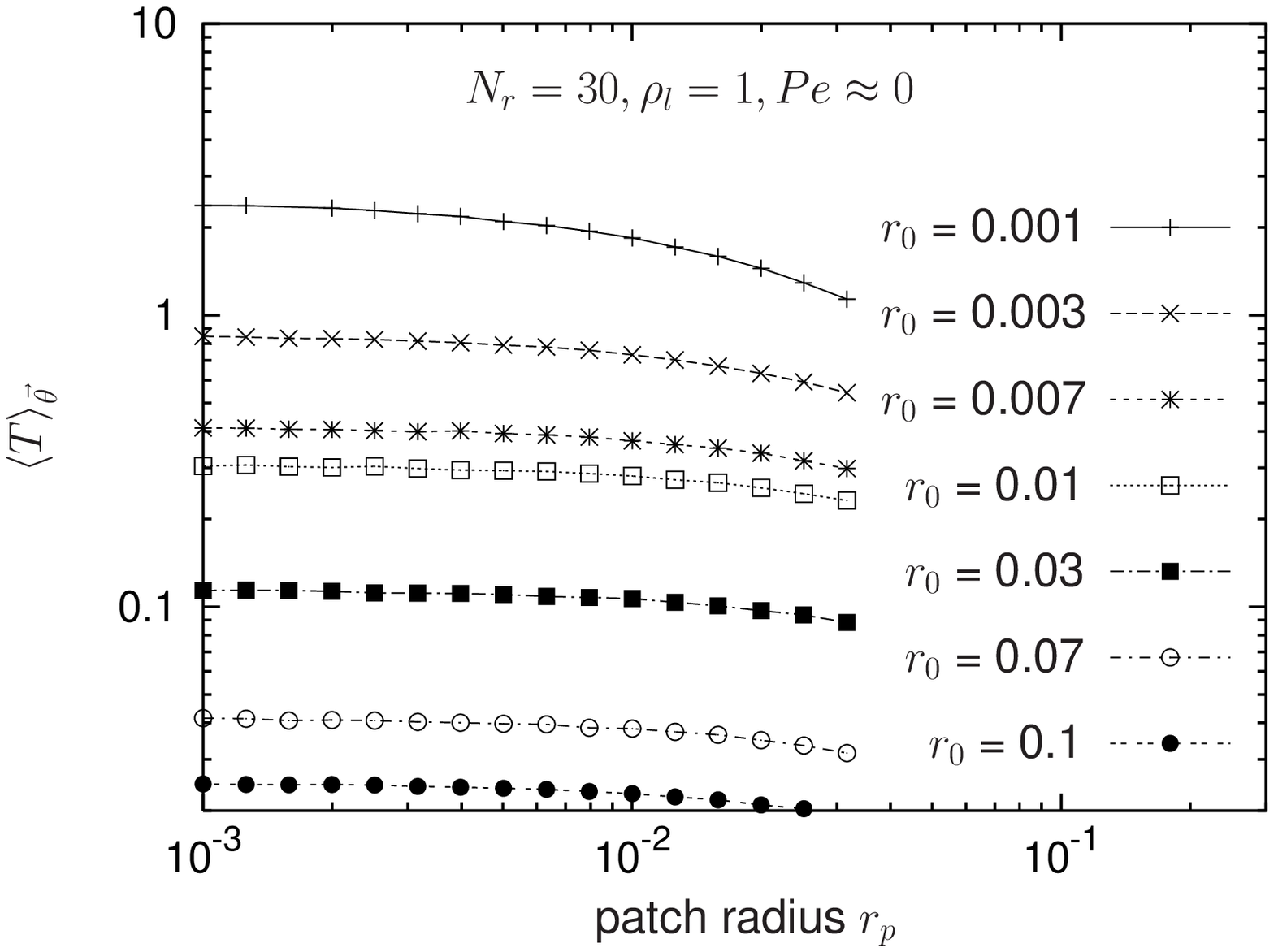}}

\fig{fig:3d:02}c
\newpage

\resizebox{.96\linewidth}{!}{\includegraphics{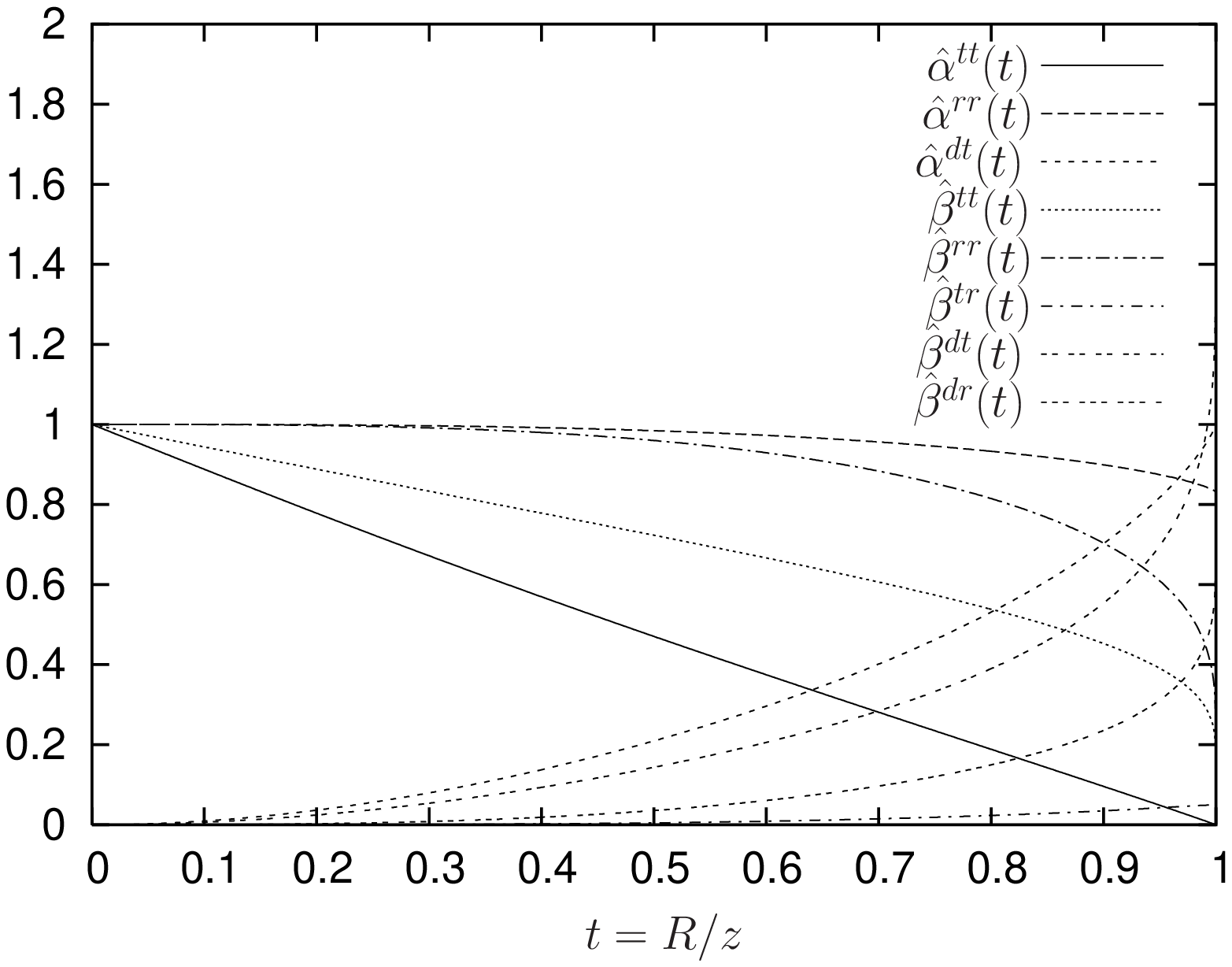}}

\fig{abbmobility}a
\newpage

\resizebox{.96\linewidth}{!}{\includegraphics{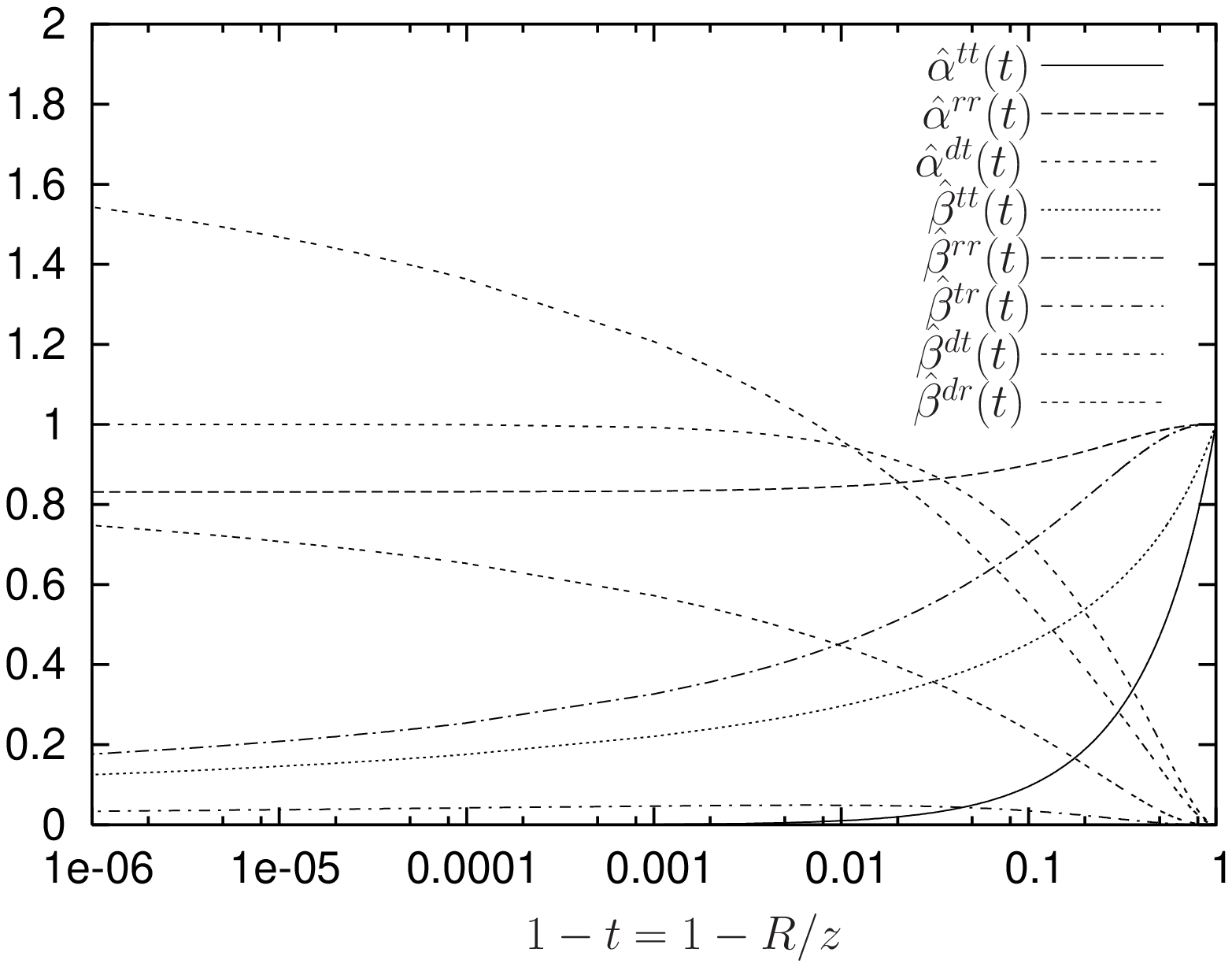}}

\fig{abbmobility}b
\newpage

\resizebox{.96\linewidth}{!}{\includegraphics{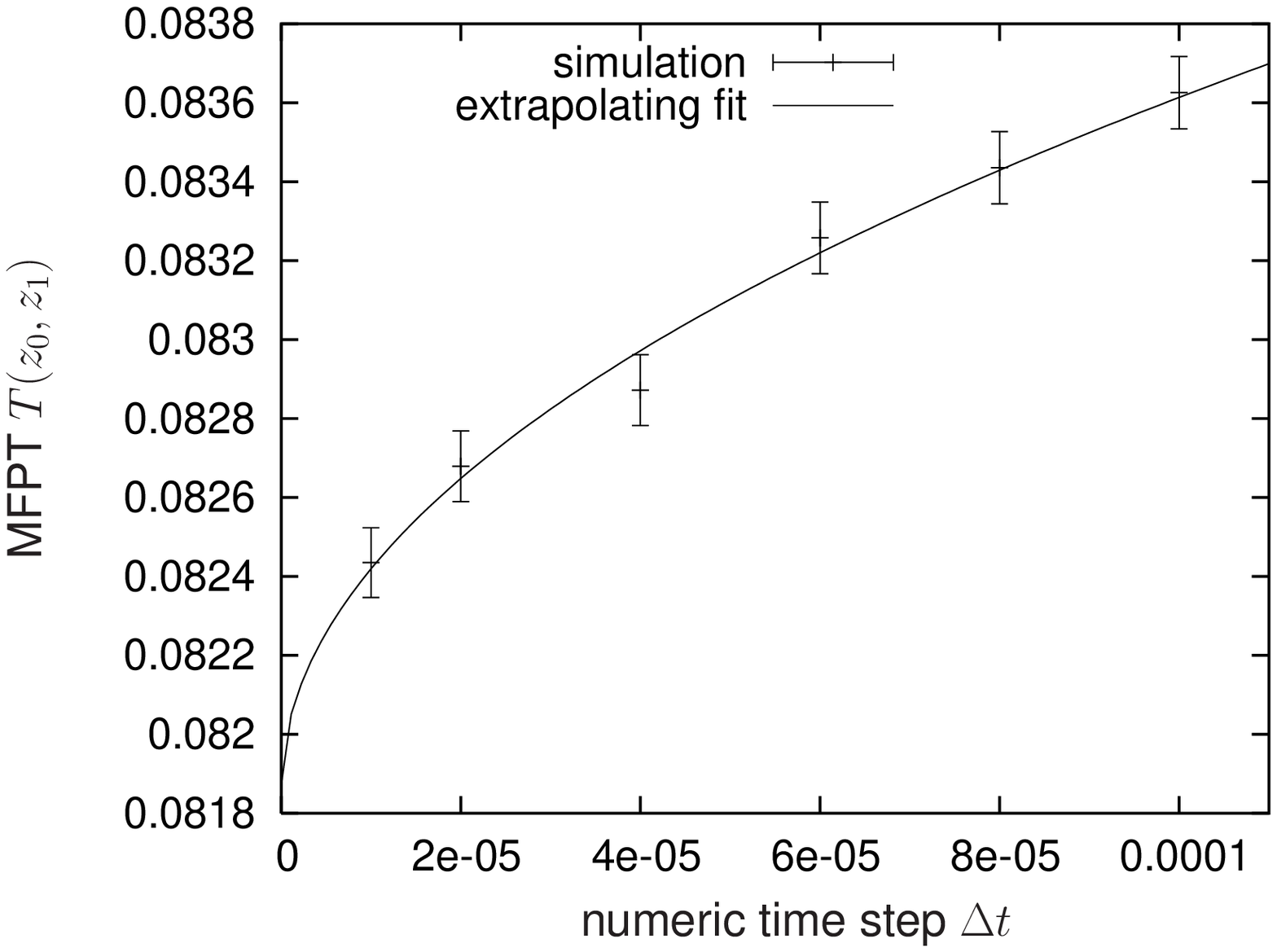}}

\fig{error}
\newpage

\end{document}